\documentstyle[aps,preprint]{revtex}
\newcommand{\be}{\begin{equation}}
\newcommand{\ee}{\end{equation}}
\newcommand{\bea}{\begin{eqnarray}}
\newcommand{\eea}{\end{eqnarray}}
\begin{document}
\bibliographystyle{prsty}
\title{Solitary Waves under the Influence of a Long-Wave Mode}
\author{Hermann Riecke}
\address{
Department of Engineering Sciences and Applied Mathematics \\
Northwestern University, Evanston, IL 60208, USA
}
\maketitle
\begin{abstract}

\vspace*{-1cm}
\baselineskip=15pt
The dynamics of solitons of the nonlinear Schr\"odinger equation under the influence of
 dissipative and dispersive perturbations is investigated. 
In particular a coupling to a 
long-wave mode is considered using extended Ginzburg-Landau equations. 
The study is motivated by the experimental observation
 of localized wave trains (`pulses') in binary-liquid convection.
These pulses have been found to drift exceedingly slowly. The perturbation analysis reveals
two distinct mechanisms which can lead to a `trapping' of the pulses by the long-wave concentration mode. They are given by the effect of the concentration mode
on the local growth rate and on the frequency of the wave. The latter, 
dispersive mechanism has not been recognized previously, despite the fact that
it dominates over the dissipative contribution within the 
perturbation theory. A second unexpected result is that 
the pulse can be accelerated by the concentration mode 
despite the reduced growth rate ahead of the pulse. The dependence of the 
pulse velocity on the Rayleigh number is discussed, and the hysteretic 
`trapping' transitions suggested by the perturbation theory 
are confirmed by numerical simulations, which also reveal 
oscillatory behavior of the pulse velocity in the vicinity of the transition. 
The derivation and reconstitution
of the extended Ginzburg-Landau equations is discussed in detail.

\vfill
\centerline{submitted to Physica D February 8, 1995}
\centerline{revised version, October 9, 1995}
\vfill

\noindent
{\bf This paper can also be obtained from patt-sol@xyz.lanl.gov via the
command {\it get 9502006}}

\end{abstract}


\section{Introduction}
The appearance of stable localized structures in spatially homogeneous extended systems
constitutes an exciting feature of various
pattern-forming systems. Although the linear instabilities arise from extended modes,
 various nonlinear mechanisms can lead to the confinement of the structure
to a small part of the system, the size of which is independent of the system size. 

In systems exhibiting steady structures the competition between structures with two
or more different
wave numbers can lead to stable domains of one wave number within
a background of a different wave number \cite{HeVi92,BrDe89,RaRi95,RaRi95a}. Such 
structures have been observed in convection experiments in narrow channels\cite{HeVi92}.
They are expected to be particularly accessible in parametrically driven surface waves in ferrofluids \cite{RaRi95b}. The stability
of domain structures is due to an oscillatory interaction between the domain walls
separating the domains \cite{RaRi95,RaRi95a}. Zig-zag structures constitute a 
two-dimensional analog of these domains \cite{BoKa90}. 

A different class of localized structures has been found in systems exhibiting a 
parity-breaking bifurcation from a steady structure to traveling waves \cite{FlSi91,GoGu91,RaCo91,WiAl92,CaCa92,RiPa92,BaMa94}. 
Instead of extended traveling waves one observes that the parity
(reflection symmetry) of the structure is broken only over a stretch of a few wavelengths.
Within that part, which is drifting through the systems, the structure is traveling,
 whereas the remaining structure is motionless. 
   Here the localization is connected with the fact that generically the 
extended traveling waves are unstable to side-band perturbations right at their onset
 \cite{CaCa92,RiPa92,BaMa94}. Remarkably, the parity-breaking bifurcation need not
be subcritical for this localization to occur \cite{CaCa92,RiPa92}. 

In this paper localized traveling waves are investigated which arise in 
the convection in binary liquid mixtures. This system has been the subject of extensive
experimental and theoretical studies. In quasi-one-dimensional set-ups
`pulses' of traveling waves have been observed 
with lengths ranging from a few up to 15 wavelengths 
\cite{KoBe88,BeKo90,NiAh90,MoFi87,Ko91}. 
In sufficiently homogeneous systems they drift
extremely slowly through the system \cite{Ko91}. Outside the pulses the fluid
 is in the conductive,
motionless state. In two-dimensional systems such pulses have also been observed but
were always found to be (very weakly)  unstable \cite{LeBo93}. A more detailed
discussion can be found in a recent review \cite{CrHo93} and in \cite{BaLu94,BaLu94a}.

Theoretically, the pulses in binary-mixture convection have posed various intriguing problems.
These are related to the mechanism of localization, the extremely slow drift of the pulses
 and their stability. Numerical simulations of the Navier-Stokes equations have 
provided quite some insight into the physics of the localized (and the extended) 
waves and in particular into the role of the concentration
field \cite{BaLu94,BaLu94a,BaLu91,LuBa92}. In order to get a detailed understanding of the mechanisms leading to the localization of 
the traveling waves and the slowing-down of the resulting pulse it is highly desirable to
study the problem also analytically. However, in particular due to the subcritical nature of 
the bifurcation to traveling waves, no rigorous asymptotic reduction of the Navier-Stokes
equation to a set of simpler evolution equations describing pulses appears available.
Nevertheless, asymptotic analysis allows the derivation of equations of the Ginzburg-Landau
type which capture and elucidate the essential mechanisms of the system. 

Various authors have shown that the dispersive character of the 
waves can lead to the localization of the waves. In the limit of strong dispersion the complex
Ginzburg-Landau equation describing
the backward Hopf bifurcation to traveling waves was considered as a dissipatively
perturbed nonlinear Schr\"odinger equation \cite{ThFa88,FaTh90,Pi87}. It was shown that the dissipative 
perturbations could select a stable solitary wave from the continuum of soliton solutions. In
the opposite limit of weak dispersion the interaction of fronts connecting the 
conductive and the convective
state were considered. It was shown that dispersion can lead to a spatially
 oscillatory contribution to
 the interaction, thus allowing stably bound pairs of fronts forming a solitary wave \cite{MaNe90,MaNe91,HaJa90,HaPo91}.

In both analytical approaches serious qualitative disagreements with the experimental observations 
remain. First,
the theoretical drift velocity of the pulses \cite{ScZi89,ScZi93,CrKi88} is found to be by a factor of 20 to 40 and more larger
than that observed in experiment \cite{Ko91,Ko94}. In particular, the theoretical pulse velocity is always in 
the same direction as the phase velocity of the waves. Experimentally, however, backward
traveling pulses are found as well \cite{Ko94}. It has been argued that this discrepancy could
be eliminated if nonlinear gradient terms were kept in the Ginzburg-Landau equation \cite{DeBr90}. 
This would, however, require that the linear group velocity be compensated for by
 higher-order correction terms, which would constitute a quite delicate balance.
  Moreover, within the Ginzburg-Landau equation
small and large pulses coexist with the large pulse being stable and the small one always
being unstable \cite{MaNe90,HaJa90}. In recent
experiments, however, only a single (stable) pulse was found for strongly negative
 separation ratio $\psi$ and two pulses for weakly negative $\psi$. 
In the latter case it was the shorter pulse which was stable 
whereas the longer pulse was unstable \cite{Ko94}. 

As pointed out in previous papers mass diffusion in liquid mixtures is extremely
slow as compared to heat diffusion \cite{Ri92}. This introduces an additional 
slow time scale 
connected to a concentration mode and 
leads to the break-down of the complex Ginzburg-Landau equations already for quite
small amplitudes of convection. Based on this observation extended Ginzburg-Landau 
equations were introduced which capture this aspect \cite{Ri92,Ri92a}.
Numerical simulations of these equations showed that the additional mode can lead to a 
considerable slow-down of the pulses \cite{Ri92}. An analysis of the interaction of fronts
\cite{HeRi94} showed that this mode introduces a new mechanism for 
localization, which is independent of dispersion. Strikingly, 
it depends strongly on the direction of propagation of the pulse \cite{HeRi95}.
Numerical simulations of the extended equations revealed that more than two pulse solutions
can coexist. In that case the longer pulse can be unstable in agreement with the experimental
results \cite{RiRa95}. 

In the present paper I investigate the extended  Ginzburg-Landau equations analytically in
detail. I consider the strongly dispersive case and study in particular the effect of the 
concentration mode on the perturbed soliton solution. In certain limits
the concentration mode can be eliminated and leads to a change of the coefficients
in the Ginzburg-Landau equation. In particular, it introduces nonlinear gradient terms. Most notably,
its contribution to the coefficients depends on the velocity of the pulse. This leads to a
strongly nonlinear dependence of the pulse velocity on the parameters.
The calculation shows that there are two different mechanisms that can lead to a slow-down
(`trapping') of the pulse. As found in numerical simulations of the
Navier-Stokes equations \cite{BaLu94} the concentration mode strongly reduces the
buoyancy of the liquid ahead of the pulse. This can - but need not - slow down
the pulse. There is in fact the possibility that the pulse is sped-up
by the concentration mode. This is due to the fact that
the variation in buoyancy also leads to a change in the wave number of the pulse which is 
directly related to its velocity. The second mechanism does not rely on a change
in the buoyancy. It arises from the change in frequency of the wave due to the concentration
mode and can also induce `trapping'. This was not anticipated from the numerical
simulations \cite{BaLu94}.

This paper is organized as follows. In sec.\ref{s:ecgl} the 
extended Ginzburg-Landau equations are presented. Their validity  and their relevance
to other systems exhibiting an interaction between short-wave and long-wave modes
is discussed in some detail. The concentration field 
generated by a pulse within this framework
is described in sec.\ref{s:conc}. The main results for the 
influence of the concentration mode on the 
pulse and its velocity are obtained in sec.\ref{s:pert} using perturbation 
theory around the soliton
of the nonlinear Schr\"odinger equation. Numerical simulations supporting the conclusions
based on the perturbation result are presented in sec.\ref{s:num}. In two appendices
a slightly more general derivation of the extended Ginzburg-Landau equation 
from the Navier-Stokes equations is presented, which leads to quantitative agreement
with the usual Ginzburg-Landau equation in a suitable limit. Their self-consistency is 
confirmed using Ward identities.

\section{The Extended Ginzburg-Landau Equations}
\label{s:ecgl}
As indicated in the introduction, the pulses of the complex Ginzburg-Landau equation differ 
 in various aspects {\it qualitatively} from those observed experimentally. 
The numerical simulations
of the Navier-Stokes equations indicate that certain modes of the concentration field  are
independent of the convective amplitude \cite{BaLu94}. 
In the complex Ginzburg-Landau equation the 
concentration field is, however, adiabatically slaved to the convective amplitude. 
The concentration field can  become an independent dynamic quantity only
if its time scales are slow, i.e. of the same order as those of the convective amplitude.
In \cite{Ri92} it was recognized that in binary-liquid
convection this is indeed the case since the mass diffusion is by a factor 100 slower than
heat diffusion. Since the Ginzburg-Landau equation is only valid as long as the time 
scales of the convective amplitude are slow as compared to all other time scales, the
slow mass diffusion 
implies that the regime of validity of the Ginzburg-Landau equation
 is limited to very small amplitudes.

To capture the additional dynamics of the concentration field I have 
considered the Navier-Stokes equations  
in an expansion in which the Lewis number, which characterizes the ratio
of mass diffusion to heat diffusion, is of the same order as the growth rate 
of the convective amplitude \cite{Ri92}. 
In the relevant limit of vanishing mass diffusion two types of critical modes arise:
the usual convective
mode $A$, which involves the stream function, the temperature and the concentration field,
and an infinite number of critical modes which are all associated with the 
concentration field alone: 
any concentration mode which is independent of the horizontal
coordinate has vanishing growth rate. Strictly speaking, this implies that in a 
weakly nonlinear theory this infinite number of modes would have to be kept.
However, since all these modes are damped
for non-zero Lewis number, it is reasonable to keep only 
those modes which are driven directly by the convective amplitude. 

For free-slip-permeable
boundary conditions it turns out that only a single mode $C$ is excited to lowest order. 
The form of the evolution equation for $A$ and $C$ can then be derived without any
detailed calculation and irrespective of the boundary conditions
by using the translation
and reflection symmetries of the system. One obtains to third order \cite{Ri92}
\bea
\partial_tA+(s+s_2C)\partial_xA&=&d\partial_x^2A+(a
+fC+f_2C^2+f_3\partial_xC)A+cA|A|^2+...,\label{e:caa}\\
\partial_tC&=&\delta \partial_x^2C-\alpha C+h_2\partial_x|A|^2+
(h_1+h_3C)|A|^2+ih_4(A^*\partial_xA-A\partial_xA^*)+....\label{e:cac}
\eea
For these boundary conditions the coefficients have been calculated explicitly from the Navier-Stokes equations \cite{Ri92a}. Due to the purely advective nature of the nonlinearity
in the Navier-Stokes equation, no nonlinear terms of the form $C^n$ arise in
(\ref{e:caa},\ref{e:cac}). Since $C$ is even under reflections in $x$ no terms like
$\partial_x C^n$ can arise.
 Note that in \cite{Ri92,Ri92a,HeRi95} the term
proportional to $h_4$ has been given incorrectly. It describes the wave-number
dependence of the mixing term $h_1|A|^2$. 
A somewhat generalized derivation of (\ref{e:caa},\ref{e:cac})
is presented in appendix \ref{s:appa}. 

For realistic boundary conditions more modes may arise. Considering a complete
set of vertical eigenfunctions $C^{(n)}(z)$, it is in general to be expected that a whole 
range of them is driven
simultaneously. However, their damping due to vertical diffusion (cf. coefficient
$\alpha$ in (\ref{e:cac})) grows quadratically with their number $n$ of nodes. 
In addition, their effect on the convective 
amplitude depends on the
strength of the relevant coupling coefficients (e.g. $f$ in (\ref{e:caa})). 
Since the coupling coefficients are proportional to the projection
 of the respective concentration mode on the convective amplitude they 
decrease rapidly with the number of nodes. Thus, high-order modes will have only a weak
effect. On that account the most relevant modes will be $C^{(0)}$ and $C^{(1)}$. 
In addition, the modes excited to leading order will be odd in the vertical coordinate,
since the  convective mode is even and the nonlinear operator is odd
(cf. (\ref{e:ns})). 
Thus, although the mean
concentration mode $C^{(0)}$ experiences no damping at all
(it is actually conserved) it may still be less relevant than $C^{(1)}$ since the latter
 is excited at a lower order. 

The above arguments suggest to study   
extended Ginzburg-Landau equations for a convective
amplitude $A$ coupled to a single, weakly damped, odd concentration mode 
$C^{(1)}\equiv C$. The importance of such an antisymmetric concentration 
mode has also become apparent in full
numerical simulations with realistic boundary conditions \cite{LuBa92,BaLu94}.
Since $C$ is
invariant under translations and reflections in the horizontal coordinate
 it can couple to all terms in the equation for
$A$ as demonstrated in (\ref{e:caa}) to third order. It is expected, however, that it
will affect  $A$ most strongly through the change of the small linear growth rate, 
which is one of the expansion parameters in the problem, and the linear 
frequency $via$ the term $fCA$. A striking effect of the concentration field on the local
buoyancy has been identified in  \cite{LuBa92,BaLu94}. In the following I 
will therefore consider the somewhat simplified equations
\bea
\partial_t A + s \partial_x A& = & d \partial_x^2 A + (a +fC) A + c|A|^2A + p|A|^4 A + g_1 A \partial_x |A|^2 + g_2 |A|^2 \partial_x A,\label{e:ecglA}\\
\partial_t C & = &\delta \partial_x^2 C - \alpha C + h_1 |A|^2+ h_2 \partial_x |A|^2.
\label{e:ecglC}
\eea
 This version seems to capture the essential mechanisms of the experimental system. 
To account for the subcritical nature of the bifurcation fifth-order 
terms as well as cubic gradient terms have been included\footnote{In \cite{Ri92} no
quintic terms had been displayed although the saturating term $p|A|^4A$ had been used in 
the numerical simulations.}. 
Note that (\ref{e:ecglA},\ref{e:ecglC}) have to be considered as reconstituted 
equations\cite{Ro85}; 
they are obtained by combining the solvability conditions arising at various orders. 
Therefore not all terms are of the same order (cf. $h_1 |A|^2$ and 
$h_2 \partial_x |A|^2$). 

In the analysis I will focus in particular on the effect of 
the gradient term $h_2 \partial_x |A|^2$
which expresses the advection of the concentration field by the traveling wave. 
As shown by Barten {\it et al.} 
the wave generates a concentration current which is antisymmetric with respect to 
the mid-plane of the convection layer \cite{LuBa92,BaLu94}.
 If the convective amplitude varies in space such a current increases the
concentration in the top half of the layer, say, and decreases it in the bottom half, 
thus generating an odd concentration mode $C$. The other gradient
 term $ih_4 (A^*\partial_x A-A\partial_x A^*c)$ will, however, be neglected; for free-slip-permeable
boundary conditions the coupling coefficients $h_1$ and $h_4$ are both of the order
of the Lewis number and therefore small. Since the term involving $h_4$ 
expresses the dependence of the mixing term $h_1|A|^2$ on the wave number it
represents a correction to a term which is already small. 

 Note that the diffusive term
$\delta \partial_x^2 C$ is not due to molecular diffusion, 
which contributes only at higher order (cf. (\ref{e:defdelta})); 
instead, it arises through the effect of large-scale variations
in the concentration field on the  local buoyancy of the fluid which
generate vorticity which in turn advects the basic (linear) concentration profile. 

It should be emphasized that the nonlinear coefficients of the equation for $A$ are not 
the same as those in the usual Ginzburg-Landau equation since the latter
 is obtained from (\ref{e:caa},\ref{e:cac})
in the limit of large Lewis number ($|a| \ll \alpha$) by eliminating $C$ adiabatically. 
This leads to additional contributions to the coefficients of (\ref{e:caa}). 
The elimination is, however, not
completely trivial. If one were to take only the terms to the order given in (\ref{e:cac})
 one would not recover the usual values of the coefficients
 (not even their limit of vanishing Lewis number,
cf. (\ref{e:defCal})). This is due to the fact that terms of higher order in the Lewis number,
which have been dropped in (\ref{e:cac}), become relevant.
In appendix \ref{s:appa} a derivation of (\ref{e:caa},\ref{e:cac}) is given in which
those higher-order terms are retained. The adiabatic elimination of $C$ from these equations
leads then to the usual coefficients in the Ginzburg-Landau equation.

The extended Ginzburg-Landau equations  (\ref{e:caa},\ref{e:cac}) 
are not only relevant for binary-mixture convection. They
describe quite generally the interaction of an unstable short-wave mode with a 
stable long-wave mode. Such an interaction arises, for instance, 
in oscillatory convection with a free surface or for waves
on the interface of two immiscible fluids in a channel. There the interface position represents the long-wave mode. The latter case has
been investigated in detail in \cite{ReRe93}. The interaction of short- and long-wave
modes is also important in waves traveling on propagating fronts. There the front
position represents the long-wave mode. In the presence of translation symmetry
its position itself is irrelevant for the dynamics; only its gradients can enter the 
equations. In contrast to the case discussed here, the relevant long-wave mode 
$V \equiv \partial_x C$ transforms
then like $V \rightarrow -V$ under reflections in $x$ \cite{Ma92a}.
In spirit, the extended Ginzburg-Landau equations (\ref{e:caa},\ref{e:cac}) 
are related to those describing patterns in the presence of
Galilean invariance \cite{CoFa85a} or coupled to a weakly damped mean flow \cite{SiZI81,Be94,DePe94}.
 
\section{The Concentration Mode Generated by a Pulse}
\label{s:conc}
To lowest order the equation for the concentration mode is a 
linear diffusion-advection equation
with damping and the concentration mode is driven by an inhomogeneous term. It is therefore
given by the integral over a suitable Green's function. 
Unfortunately, this integral appears to
have no simple closed-form solution. In this section two limiting cases are discussed. 

In order to focus on solutions with a steady envelope the extended Ginzburg-Landau equations are written in a frame moving with the pulse,
\bea
\partial_t A + (s-V) \partial_x A& = & d \partial_x^2 A + (a +fC) A + c|A|^2A + p|A|^4 A +
g_1 A \partial_x |A|^2 + g_2 |A|^2 \partial_x A \\
\partial_t C - V \partial_x C& = &\delta \partial_x^2 C - \alpha C + h_1 |A|^2+ h_2 \partial_x |A|^2 \label{e:ecglCv}
\eea
For steady solutions of (\ref{e:ecglCv}) the Green's function $G(x,x')$ is given by
\be
G(x,x')=\frac{1}{\sqrt{V^2+4\alpha \delta}} \left(e^{-k_1 (x-x')} \Theta(x-x') + e^{k_2(x-x')}\Theta(-(x-x'))\right)
\ee
with
\be
k_{1,2}=\frac{1}{2\delta}\left(\sqrt{V^2+4\alpha \delta} \pm V\right)
\ee
and $\Theta(x)$ denoting the Heaviside function.  The concentration mode can therefore be written
as
\bea
C(x)&=&\int_{-\infty}^\infty G(x,x') \left(h_1 |A(x')|^2+h_2\partial_x |A(x')|^2 \right)\,dx'
\\ \nonumber
&=&\frac{1}{\sqrt{V^2+\alpha \delta}}\int_0^\infty (h_1+h_2 k_2) e^{-k_2 x'} |A(x+x')|^2
+(h_1-h_2 k_1) k_1 e^{-k_1 x'} |A(x-x')|^2 dx'.
\eea
For the perturbation calculation in sec.\ref{s:pert} this integral will have to be evaluated for $|A(x)| =\lambda {\rm sech} (\lambda x) + h.o.t.$ 
Since I was not able to obtain a closed form for that integral I discuss two limiting cases.

For large $\alpha$ the decay rates $k_i$ become large and it is sufficient to expand
$|A(x-x')|^2$ around $x$. One then obtains 
\bea
C(x)&=&\frac{1}{\sqrt{V^2+4\alpha \delta}} \sum_{n=0}^\infty \partial_x^n |A(x)|^2 
\left((h_1+h_2k_2) \frac{1}{k_2^{n+1}} -(h_1-h_2k_2) \frac{1}{(-k_1)^{n+1}}
\right)  \nonumber\\
&=&\frac{h_1}{\alpha} |A|^2 + \left(\frac{h_1V}{\alpha^2}+\frac{h_2}{\alpha}\right)
 \partial_x |A|^2 \nonumber \\
& &+\left( \frac{h_2V}{\alpha^2}+\frac{h_1}{\alpha} 
(\frac{V^2}{\alpha^2}+\frac{\delta}{\alpha})\right)\partial_x^2 |A|^2 + 
\frac{h_2}{\alpha} \left(\frac{V^2}{\alpha^2}+\frac{\delta}{\alpha}\right) \partial_x^3 |A|^2 
+ O(\frac{1}{\alpha^4}). \label{e:defCal}
\eea
Thus, the concentration mode changes the coefficients of the
nonlinear gradient terms in the equation
for the convective amplitude $A$ (or introduces such terms if they should have been absent). 
It is noteworthy that the contributions to these terms depend on the velocity of the pulse. 
In this limit, however, the concentration mode can follow adiabatically the
convective amplitude and one obtains essentially the usual Ginzburg-Landau equation.

Another, more interesting limit is that of large velocity (or equivalently 
small $\alpha$ and small $\delta$). In this case the time scale
set by the velocity of the pulse is short compared to the relaxation time 
of the concentration mode. The concentration mode can therefore not follow adiabatically.
As a consequence the spatial decay rate corresponding
to the decay ahead of the pulse 
is large whereas that describing the decay of the concentration mode behind the pulse
is small. For $V>0$ the rates are given by
\be 
k_1 = \frac{V}{\delta} \left(1 + \frac{\alpha \delta}{V^2} + O(V^{-4})\right), \ \ \ \  k_2=\frac{\alpha}{V}+O(V^{-3}).
\ee
The slow decay behind the pulse allows an expansion of the exponential and one
obtains
\bea
C(x)&=\frac{1}{\sqrt{V^2+4\alpha \delta}} &\left\{ (h_1-h_2k_2) \sum_{n=0}^\infty 
\frac{-1}{(-k_1)^n} \partial_x^n |A(x)|^2+\right. \nonumber \\
& & \left.(h_1+h_2k_2)\sum_{n=0}^\infty  (-k_2)^n \int_0^\infty x'^n|A(x+x')|^2\,dx' 
 \right\}.
\eea
For $V<0$ one has
\be
k_1=-\frac{\alpha}{V}+O(V^{-3}),\ \ \ \  k_2=-\frac{V}{\delta}\left(1+\frac{\alpha \delta}{V^2}+O(V^{-4})\right)
\ee
and
\bea
C(x)&=\frac{1}{\sqrt{V^2+4\alpha \delta}}& \left\{
(h_1-h_2k_1) \sum_{n=0}^\infty (-k_1)^n \int_0^\infty x'^n|A(x-x')|^2\,dx'+
\right.\nonumber \\
& &\left. (h_1+h_2k_2) \sum_{n=0}^\infty \frac{1}{(k_2)^{n+1}}\partial_x^n |A(x)|^2 
\right\}. 
\label{e:Cexp}
\eea
After inserting the expansions for $k_i$ into (\ref{e:Cexp})
one obtains for the concentration mode 
\bea
C(x)&=&\left(\frac{h_1\delta}{V^2}-\frac{h_2}{V}\right) |A(x)|^2 +
 \frac{h_2 \delta }{V^2} \partial_x |A(x)|^2 + 
\left( \frac{h_2\alpha}{V^2} +\frac{h_1}{V} \right) S \int_0^\infty |A(x + Sx')|^2\,dx' \nonumber \\
& &-\frac{h_1 \alpha}{V^2}  \int_0^\infty x' |A(x+ Sx')|^2 \,dx'  + O(\frac{1}{V^3})
\label{e:defC}
\eea
with $S=|V|/V$. This expansion is not valid uniformly in space since 
the gradients in $C(x)$ become small far behind the soliton 
without $C(x)$ itself being small. The damping
term therefore becomes important and ensures that $C(x)$ goes to 0 sufficiently
far behind the pulse. This can
be taken care of by a matching procedure. For the perturbation calculation below this
is, however, not necessary since the soliton amplitude decays exponentially there.
Since the envelope of the soliton of the  unperturbed nonlinear Schr\"odinger equation 
is even, 
the odd contributions to $C$ will affect the velocity of the pulse. In particular, for 
$h_1 =0$ one 
expects a relationship between the group velocity $s$ and the pulse velocity $V$ of
the form
\be
V = s + \frac{h_2}{V^2} K + O(V^{-4}), \label{e:vs}
\ee
where the coefficient $K$ depends on the amplitude of the pulse. This is confirmed by the
perturbation analysis below in which $K$ is calculated in detail. Thus, in 
the presence of the concentration mode the pulse velocity is a strongly nonlinear function 
of the linear group velocity $s$ and of the other coefficients of the extended 
Ginzburg-Landau equation. Eq.(\ref{e:vs}) suggests even
the possibility of a multi-valued connection between $V$ and $s$ and a hysteretic transition
between the different solution branches. This rich behavior 
is to be contrasted with that described by the complex Ginzburg-Landau
equation alone for which the pulse velocity is strictly linear in $s$ even if nonlinear gradient
terms are included \cite{DeBr90}.

\section{The Soliton Evolution Equations}
\label{s:pert}
\subsection{Perturbed Nonlinear Schr\"odinger Equation}
In this section the evolution equations for the solitary wave are derived by
considering it as a
perturbed soliton. The nonlinear Schr\"odinger equation has four continuous symmetries:
translations in space, phase shifts,
scaling of the amplitude and Galileian invariance. 
The soliton solutions therefore form 4-parameter families of solutions
characterized by their location $x_0$, phase $\phi_0$, amplitude $\lambda$ and wave number
$q$. Under small dissipative
and dispersive perturbations, as they are introduced by
the extended Ginzburg-Landau equations (\ref{e:ecglA},\ref{e:ecglC}), the dynamics is predominantly within
one of these families and can be characterized by the slow 
evolution of $x_0(t)$, $\phi_0(t)$,
$\lambda(t)$ and $q(t)$. The corresponding evolution equations are obtained from 
solvability conditions arising in the perturbation expansion. In the nonlinear Schr\"odinger
equation these are due not only to proper but also to generalized zero-eigenvectors 
of the linearized operator in question \cite{We85,ElMe89}. 

Thus, I consider the perturbed nonlinear Schr\"odinger equation 
\cite{ThFa88,FaTh90,Pi87,We85,ElMe89,Ma87}
\be
i\partial_t A+i (s-V) \partial_x A + \frac{1}{2} \partial_x^2 A + |A|^2A= \epsilon P, \label{e:pnls}
\ee
where $P$ contains the remaining terms of the extended Ginzburg-Landau equations including
the contribution from the concentration mode through eq.(\ref{e:defC}). Since the 
explicit expressions for the concentration mode were only obtained in a frame in which
the soliton is at rest, eq.(\ref{e:pnls}) has been transformed to a frame moving with
velocity $V$ and it is this velocity that needs to be determined. 
The amplitude $A$ is expanded
around the single-soliton solution,
\be
A(x,t)=\left( A_0(\Theta)+ \epsilon A_1(\Theta,T) +...\right) 
e^{i q(T) \theta + i \phi}, \label{e:Aexp}
\ee
with 
\be
A_0 = \lambda(T) {\rm sech} (\Theta), \ \ \ \  \Theta=\lambda(T) \theta
\ee
and  
\be
\partial_t \theta = -v(T),\ \ \ \  \partial_x \theta = 1,
\ \ \ \  \partial_t \phi = \omega(T), \ \ \ \  \partial_x \phi = 0.
\ee
The quantities characterizing the soliton within the family are allowed to vary slowly in time
in order to eliminate the secular terms arising in the perturbation expansion.
In addition, they - as well as $V$ - are expanded in $\epsilon$,
\bea
\lambda = \lambda_0 + \epsilon \lambda_1 + ..., \ \ \ \  
q = q_0 + \epsilon q_1+...,\ \ \ \ v = v_0 + \epsilon v_1+..., \\
\omega = \omega_0 + \epsilon \omega_1 + ...,\ \ \ \ V = V_0 + \epsilon V_1 + ... 
\label{e:parexp}
\eea
Note that the perturbation $P$ can also change the width of the soliton. Therefore
a long-wave theory, in which $\Theta$ is a slowly varying function of $x$ and $t$ is not
sufficient \cite{Gr79}. In the present ansatz any 
time-dependence of $\lambda$ and $q$ leads to
diverging time derivatives of the phase for large values of $\theta$. This poses, 
however, no problem here since the (bright) soliton goes to 0 exponentially fast there.
Inserting the ansatz (\ref{e:Aexp})-(\ref{e:parexp}) 
into (\ref{e:pnls}) yields at lowest order
\be
v_0= s-V_0+q_0, \ \ \ \omega_0 = \frac{1}{2} (\lambda_0^2 + q_0^2). \label{e:defv0}
\ee
Thus, to this order the amplitude and the wave number as well as the velocity
and the frequency of the pulse are undetermined. At $O(\epsilon)$ one obtains
\be
i \partial_t A_1+ {\cal L} A_1 \equiv i \partial_t A_1 - \frac{1}{2} \lambda_0^2 A_1+ \frac{1}{2} \lambda_0^2 \partial_\Theta^2 A_1 + A_0^2 (2 A_1 + A_1^*) = I, 
\label{e:linear}
\ee
where the inhomogeneity $I$ contains in addition to $P$
 the contributions from the derivatives 
of the slowly varying quantities. The operator $i{\cal L}$ is singular. It has two proper 
zero-eigenvectors $iA_0$ and $\partial_\Theta A_0$, which arise from the translation and phase shift symmetry of the unperturbed
nonlinear Schr\"odinger equation. In addition, $i{\cal L}$ has two generalized
eigenvectors $i\Theta A_0$ and $\Theta \partial_\Theta A_0 + A_0$ \cite{We85,ElMe89},
\be
i{\cal L} \, \left\{i\Theta A_0 \right\}= - \lambda_0^2 \partial_\Theta A_0,\ \ \ \ 
i{\cal L} \, \left\{\Theta \partial_\Theta A_0 + A_0 \right\}= i\lambda_0^2A_0,
\ee
which arise from the scale and the Galileian invariance. In general, $A_1$ will
contain secular terms. Weinstein has shown \cite{We85} that $A_1$ remains bounded
for times of $O(1/\epsilon)$ if $I$ is orthogonal to the 
generalized null-space of $i{\cal L}$, in which case $A_1$ is also orthogonal to that space.
To project $iI$ onto that null-space an appropriate scalar product is defined by
\be
(A,B)= Re \left( \int_{-\infty}^\infty A^*(\Theta) B(\Theta) d\Theta \right).
\ee
It corresponds to the usual scalar product on ${\cal R}^2$ once the complex quantities are separated in real and imaginary parts. 
Using the appropriate left eigenvectors, the four solvability conditions are given by
\bea
Re \int_{-\infty}^\infty i A_0 I \,d\Theta =0, \\
Re \int_{-\infty}^\infty \partial_\Theta A_0 I \,d\Theta = 0, \\
Re \int_{-\infty}^\infty i \Theta A_0 I \,d\Theta = 0,\\
Re \int_{-\infty}^\infty (A_0 +\Theta \partial_\Theta A_0) I \,d\Theta = 0.
\eea
Inserting the expansion (\ref{e:Aexp}) one obtains
\bea
\,{\frac {d}{dT}}\lambda_0&=&\int _{-\infty }^{\infty }\!{
\rm sech}(\Theta)P_i\,{d\Theta}\label{e:lambda0}\\
\,{\frac {d}{dT}}q_0&=&-\int _{-\infty }^{\infty }\!{\rm sech}(\Theta)\tanh(\Theta)P_r 
\,{d\Theta}\label{e:q0g}\\
v_1&=&q_1-V_1+{\frac {1}{\lambda_0^{2}}\int _{-\infty }^{\infty }\!\Theta
\,{\rm sech}(\Theta)P_i\,{d\Theta} \label{e:v1g}
}\\
\omega_1&=&\lambda_0 \lambda_1+q_0 (V_1 +v_1) \nonumber \\
& &-\frac {1}{\lambda_0}\int _{-\infty }^{\infty }\!{\rm sech}(\Theta)P_r \,{d\Theta}+ 
\frac {1}{\lambda_0}\int _{-\infty }^{\infty }\!\Theta\,{\rm sech}(
\Theta)\tanh(\Theta)P_r\,{d\Theta} \label{e:omega1}
\eea
with $P=P_r + i P_i$. Eqs.(\ref{e:lambda0},\ref{e:q0g}) yield evolution equations for the lowest-order 
contributions to the amplitude and to 
the wave number. This determines also the velocity and frequency
to lowest order (cf. (\ref{e:defv0})). 
The quantity of most interest in the present context is the effect of  
 the concentration mode on the velocity of the soliton. To lowest order it is
determined by the group velocity $s$ and the wave number $q_0$, $V_0 =s +q_0$. 
The relevant question is therefore how the perturbations affect the wave number. 
It turns out that spatial variations in the growth rate due to the
concentration mode, which were also identified
 in the numerical simulations of the Navier-Stokes equations, 
lead to $q_0 =0$ independent of their strength.
Therefore, no change in velocity arises at this order.
One then has to determine $V_1$ which in turn depends on $q_1$ (cf. 
(\ref{e:v1g})). This contribution to the wave number is, however,
not determined until $O(\epsilon^2)$. Spatial variations in the frequency
 on the other hand affect the wave number already at lower order.

In order to go to $O(\epsilon^2)$ one needs to determine $A_1$. This can be done by
variation of constants using the homogeneous solutions of the real and the imaginary part of
(\ref{e:linear}), respectively,
\bea
A_{1r}^{(h)}&=&{\rm sech}(\Theta)\tanh(\Theta),\\
A_{2r}^{(h)}&=&\Theta\,{\rm sech}(
\Theta)\tanh(\Theta)+\frac {1}3 \cosh(\Theta)-{\rm sech}(\Theta),\\
A_{1i}^{(h)}&=&{\rm sech}(\Theta),\\
A_{2i}^{(h)}&=&\Theta\,{\rm sech}(\Theta)+\sinh(\Theta).
\eea
The general solution is then given by
\bea
A_1=(F_{1r}+G_{1r}) A_{1r}^{(h)} + (F_{2r}+G_{2r}) A_{2r}^{(h)} + \nonumber \\
i \left((F_{1i}+G_{1i}) A_{1i}^{(h)} + (F_{2i}+G_{2i}) A_{2i}^{(h)} \right) \label{e:defA1gen}
\eea
with
\bea
F_{1r}=\frac3 {\lambda_0^2} \int  I_r A_{2r}^{(h)} \, d{\Theta},\\
F_{2r}=\frac{-3}{\lambda_0^2} \int  I_r A_{1r}^{(h)}\, d{\Theta},\\
F_{1i}=\frac{1}{\lambda_0^2} \int I_i A_{2i}^{(h)}\, d{\Theta},\\
F_{2i}=\frac{-1}{\lambda_0^2} \int I_i A_{1i}^{(h)}\, d{\Theta},
\eea
and $I=I_r+iI_i$. The coefficients $G_2$ allow the elimination of certain exponential 
divergences of $A_1$  for large $|\Theta|$. The solvability conditions arising from the
projection onto the proper zero-eigenvectors (\ref{e:lambda0},\ref{e:q0g}) eliminate the
other exponential divergences. The contributions from the coefficients $G_1$ lead only
to a fixed shift of the position and of the phase of the soliton. They can therefore be
set to 0. After inserting $A_1$ into (\ref{e:pnls}) the required equations for
$\lambda_1$ and $q_1$ are obtained by applying the solvability conditions 
at $O(\epsilon^2)$.

\subsection{First-Order Evolution Equations: Effect of $C$ on the Frequency}
\label{s:ord1}
The extended Ginzburg-Landau equations (\ref{e:ecglA},\ref{e:ecglC}) lead to the perturbation terms
\be
P = i (d_r \partial_x^2 A +a_r A + c |A|^2 A + p |A|^4A + g_1 A \partial_x |A|^2 + g_2 |A|^2 \partial_x A + f A C).
\ee
Here $C$ is given for $\alpha$ and $\delta$ small as compared to $V$ by
\bea
C &=& \left(-\frac{h_2}{V}+ \frac{h_1 \delta}{V^2} \right)|A|^2 + \frac{h_2 \delta \lambda}{V^2} \partial_\Theta |A|^2 + \nonumber \\
& &  \left( \frac{\alpha h_2}{V^2 \lambda}+\frac{h_1}{V \lambda} \right) 
\int_{ \Theta}^{S \Lambda\infty} |A(\Theta ')|^2 d \Theta ' +
\frac{h_1 \alpha}{V^2 \lambda^2} \int_{ \Theta}^{ S \Lambda \infty}
\, \left( \Theta - \Theta '\right) |A(\Theta ')|^2\, d \Theta '
\eea
with $\Lambda = |\lambda|/\lambda$. 
In view of other possible applications of this perturbation expansion 
a slightly more general perturbation is used
\bea
P& =& i (d_r \partial_x^2 A +a_r A + c |A|^2 A + p |A|^4A + (\rho_1 + i\sigma_1) A \partial_x |A|^2 +
(\rho_2 + i\sigma_2) |A|^2 \partial_x A +\nonumber \\
& &  (\rho_0+i \sigma_0) A \frac{1}{\lambda} \int_{  \Theta}^{S \Lambda \infty} |A(\Theta ')|^2 d \Theta '
 + \nonumber \\
& &(\rho_{-1}+i \sigma_{-1}) A \frac{1}{\lambda^2} \int_{  \Theta}^{ S \Lambda \infty}
\, \left( \Theta - \Theta '\right) |A(\Theta ')|^2\, d \Theta '.
\eea
Insertion of this perturbation in eqs.(\ref{e:lambda0})-(\ref{e:omega1}) leads to
\bea 
{\frac {d}{dT}}\lambda_0& =&{\frac {16\,\lambda_0 ^{5}p_{{
r}}}{15}}+\left ({\frac {4\,c_r}3 }-{\frac {2\,d_r}3 }-{
\frac {4\,\sigma_2 q_0 }3 }\right )\lambda_0 ^3 + 2 S \Lambda 
\,\lambda_0 ^{2}\rho_{{0}}+\left (2\,a_r-2\,d_r\,q_0^{2}-2\,\rho_{{-1}}\right )\lambda_0 , \label{e:la0a}\\
{\frac {d}{dT}}q_0 &=&S \Lambda \lambda_0 \sigma_{{-1}}-{\frac {4\,
d_r\,\lambda_0 ^{2}q_0 }3 }-{\frac {2\,\lambda_0^{2}\sigma_{{0}}}3 }-\left ({\frac {4\,\sigma_2 }{15
}}+{\frac {8\,\sigma_{{1}}}{15}}\right )\lambda_0 ^{4},\label{e:q0a}\\
v_1 &=&q_1-V_{{1}}-\left ({\frac {\rho_2 }3 }+{\frac {2\,\rho_{{1}}}3 }
\right )\lambda_0 ^{2}-\rho_{{0}}+ \Lambda{\frac {
\rho_{{-1}}\pi^2}{6\,\lambda_0 }},\label{e:v1a}\\
\omega_1 &=&{\frac {8\,\lambda_0 ^{4}p_i}{9}}+
\left (c_i+\rho_2 q_0 \right )\lambda_0 ^{2}+
\left (\lambda_1 + \Lambda S\sigma_{{0}}\right )\lambda_0 +q_0 (v_1+V_{{1}})-
{\frac {\sigma_{{-1}}}{2}}.
\label{e:om1a}
\eea
Eqs.(\ref{e:la0a},\ref{e:q0a}) constitute two coupled nonlinear evolution equations for the
amplitude and the wave number of the pulse. 
This could in general lead to interesting dynamics. The quadratic term 
$s S \Lambda \lambda_0^2 \rho_0$ in (\ref{e:la0a}) indicates that the subcritical 
pitchfork bifurcation is perturbed to a transcritical bifurcation.

Here I focus on a discussion of the influence of the concentration mode. 
On the one hand this restricts the choice of the coefficients  
$\sigma_j=f_i\rho_j$, $j=-1,...2$. On the other hand, the coefficients depend on the
velocity $V$ of the pulse (cf. (\ref{e:defC})) which in turn depends
on the wave number. This introduces additional complexity. Explicitly, one obtains
\bea
{\frac {d}{dT}}\lambda_0 &=&{\frac {16\,p_{{r}}\lambda_0^{5}}{15}}+\frac{4}{3}\left (\left (-{\frac {h_{{2}}}{V_{{0}}}}+{\frac {h_{{1}}\delta}{V_{{0}}^{2}}}\right )f_{{r}}+c_{{r}}-
{\frac {1}{2}}d_{{r}}-g_{{2,i}}q_0\right )\lambda_0^{3} \nonumber \\
& &+2\,\Lambda\,S\left ({\frac {\alpha\,h_{{2}}}{V_{{0}}^{2}}}+{\frac {h_{{1}}}
{V_{{0}}}}\right )f_{{r}}\lambda_0^{2}+2\,\left (a_{{r}}-d_{{r}}q_0^{2}-{\frac {f_{
{r}}h_{{1}}\alpha}{V_{{0}}^{2}}}\right )\lambda_0+O(V_0^{-3}), \label{e:la0}\\
{\frac {d}{dT}}q_0&=&-\frac{4}{15}\left ({\frac {2\,h_{{2}}
\delta}{V_{{0}}^{2}}}f_{{i}}+g_{{2,i}}+2\,g_{{1,i}}\right )\lambda_0^{4}-\nonumber \\
& &\frac{2}{3}\left (\left ({\frac {h_{{1}}}{V_{{0}}}}+
{\frac{\alpha\,h_{{2}}}{V_{{0}}^{2}}}\right )
f_{{i}}+2\,d_{{r}}q_0\right )\lambda_0^{2}+{\frac {\lambda_0f_{{i}}h_{{1}}
\alpha\,\Lambda\,S}{V_{{0}}^{2}}} +O(V_0^{-3}), \label{e:q0}\\
v_1&=&q_1-V_{{1}}-\left (\frac{1}{V_0}-{\frac {\Lambda\,S\alpha\,\pi ^{2}}{6
\,\lambda_0V_{{0}}^{2}}}\right )f_{{r}}h_{{1}}-{\frac {\left (2\,\lambda_0^{2}
\delta+3\,\alpha\right )f_{{r}}h_{{2}}}{3\,V_{{0}}^{2}}}-\nonumber \\
& &\frac{1}{3}\left (2\,g_{{1,r}}+g_{{2,r}}\right )\lambda_0^{2} +O(V_0^{-3}),
\label{e:v1}\\
\omega_1&=&\left ({\frac {\lambda_0\Lambda\,S}{V_{{0}}}}+{\frac {2\,\lambda_0^{2}\delta
-\alpha}{2\,V_{{0}}^{2}}}\right )f_{{i}}h_{{1}}+
\left (-{\frac {\lambda_0^{2}}{V_{{0}}}}+{\frac {\lambda_0\Lambda\,S\alpha}{V_{{0}}^{2
}}}\right )f_{{i}}h_{{2}}+\nonumber \\
& & q_0(v_1+V_{{1}})+g_{{2,r}}\lambda_0^{2}q_0+\lambda_0\lambda_1+{\frac {8\,\lambda_0^{4}p_{{i}}}{9}}+O(V_0^{-3}). \label{e:om1}
\eea
Eqs.(\ref{e:la0},\ref{e:q0}) show that $\lambda_0$ and $q_0$ and therefore also $V_0$ are $O(1)$-quantities and
can vary over an $O(1)$-range. The smallness of the perturbations expresses itself only 
in the slowness of the dynamics. Note, however, that the solvability conditions 
are only valid in a frame of reference in which the pulse is steady, 
i.e. $v=0$, ${\frac {d}{dT}}\lambda=0$
and  ${\frac {d}{dT}}q=0$. This limitation arises from the expansion 
for the concentration mode (\ref{e:defC}). 

Since $q_0=V_0-s$ eq.(\ref{e:q0}) 
can be viewed as
an equation for $V_0$. Solving for the group velocity $s$ one obtains
\bea
s=V_0+\frac{f_i h_1}{2\,d_r} \left( \frac{1}{ V_0}-\frac{3}{2}\frac{S \Lambda \alpha}{\lambda_0 V_0^2}\right) + f_i h_2 \frac{1}{V_0^2d_r}\left( \frac{1}{2}\alpha +
\frac{2}{5} \lambda_0^2 \delta \right) + \nonumber \\
\frac{1}{5 d_r} (2g_{1i}+g_{2i}) \lambda_0^2
+O(V_0^{-3}). \label{e:defsecgl}
\eea
Thus, the perturbation terms arising from the dynamical equation for $C$ 
lead to a strongly nonlinear connection
between $s$ and the pulse velocity $V_0$ (\footnote{Note that to leading 
order in $1/V_0$ the amplitude $\lambda_0$ is independent of $s$ and $V_0$.}). This is to 
be contrasted with the result
one would obtain from a complex Ginzburg-Landau equation alone in which the
coefficients of the nonlinear gradient terms are fixed \cite{DeBr90},
\be
s=V_0 -{\frac {3\, S \Lambda \sigma_{-1}}{4\,\lambda_0 d_r}}+{
\frac {\sigma_0 }{2\,d_r}}+\frac{1}{5}\frac{\lambda_0^{2}}{d_r}
(\sigma_2+2\,\sigma_1).\label{e:defscgl}
\ee
In (\ref{e:defscgl}) the perturbation terms lead only to a fixed shift in the 
velocity (as do the terms involving $g_j$ in (\ref{e:defsecgl})). 
Although the shift depends on the amplitude $\lambda_0$ it does not give a 
natural explanation why $V_0$ should be close to 0 over a range of parameters. 

\begin{figure}[htb]
\begin{picture}(420,240)(0,0)
\put(0,-45) {\includegraphics{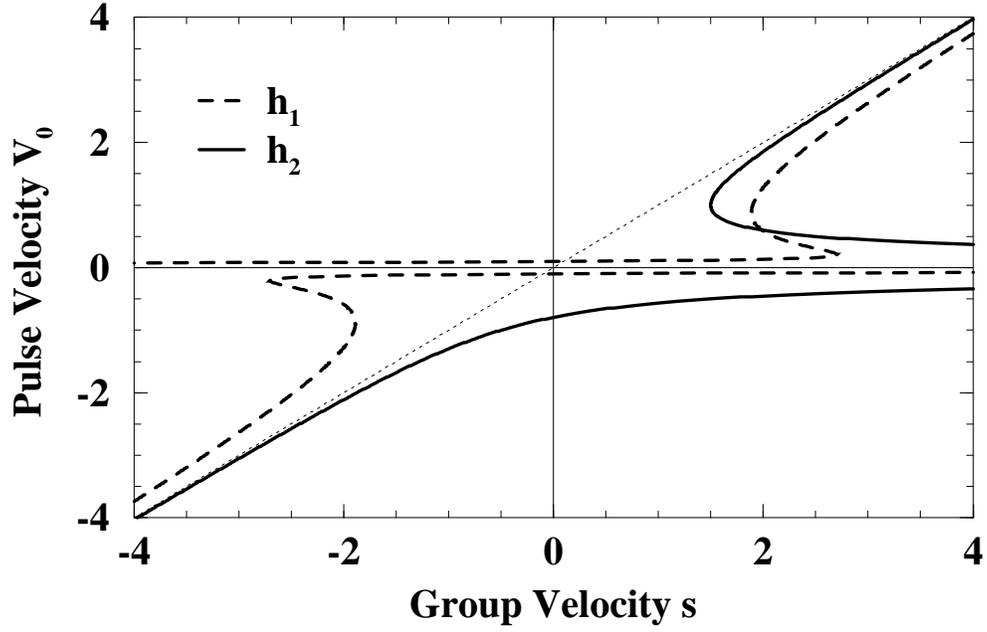}}
\end{picture}
\caption{Sketch of the contributions to the pulse velocity $V_0$ from 
the $h_1$- and the $h_2$-term (cf. (\protect{\ref{e:defsecgl}})).
\protect{\label{f:vsh12}}
}
\end{figure}

By contrast, both the terms proportional to $h_1$ and to $h_2$ 
in (\ref{e:defsecgl}) are suggestive of a range of parameters with small $V_0$. This is
indicated in fig.\ref{f:vsh12} where the contributions to $V_0 (s)$  from
the $h_1$-term (dashed line) and from the $h_2$-term (solid line) are sketched separately 
for $f_i h_1 >0$ and $f_i h_2 >0$ \cite{Ri92a}. Of course, the singularities 
at $V_0=0$
are unphysical and arise from a break-down of the expansion (\ref{e:defC}) for small $V_0$.
Strictly speaking, even the saddle-node bifurcations suggested by 
fig.\ref{f:vsh12} are beyond the present expansion. 
In that regime the diffusion and the damping term in (\ref{e:ecglCv}) come into play. 
It is expected that they will lead to a smooth connection of
the two branches of $V_0 (s)$  and 
thus to a whole range of parameters (with $s>0$) in which $V_0$ is small. 
This is confirmed by the numerical simulations discussed in 
Sec.\ref{s:num} below. 
An analytical description that removes the singularity would require a better approximation
of the concentration mode than that given by (\ref{e:defC}).
Obviously, for negative values of $f_i h_1$ and $f_i h_2$ the 
concentration mode leads to 
a speed-up of the pulse (for $s>0$). 

For free-slip-permeable boundary conditions it is found that
 $f_i h_2 >0$ and $f_i h_1 >0$ (cf.~(\ref{e:deff},\ref{e:defh2})). 
Thus, within that
approximation the extended Ginzburg-Landau equations yield a slow-down of the pulse,
as is observed in experiments. For realistic boundary conditions the coefficients in eqs.(\ref{e:ecglA},\ref{e:ecglC})
are likely to be different\footnote{Note that the nonlinear coefficients in (\ref{e:ecglA})
differ from those of the complex Ginzburg-Landau equation derived in \cite{ScZi89} 
(cf.~appendix \ref{s:appa}).}.
However, to leading order in the expansion in $1/V_0$ only the sign of the two products 
$f_i h_1$ and $f_i h_2$ is relevant to determine whether the pulse is slowed down. 
The other coefficients determine only the strength
of the effect and whether the pulse velocity shows hysteresis. 
Thus, the mechanism is quite robust and is likely to be relevant in the experimental 
system. It arises from the effect of the
concentration mode on the local frequency of the wave through the term $f_i C A$. 
The spatial variation of the concentration mode implies a spatial variation of the frequency
which leads to a differential phase winding. Since the gradient term 
$h_2 \partial_x |A|^2$ as well as the advective term
$V \partial_x C$ induce an odd contribution to $C$, the phase winding results in a change of 
the average wave number $q_0$ and as a consequence in a change in velocity.

Through the wave number the perturbation induces also a frequency shift  
$\omega_0 = \lambda_0^2/2 + q_0^2/2-q_0v_0$ (as compared to the Hopf frequency)
 of the pulse in the lab frame. It is worth noting that this change in frequency
differs from the frequency shift $\omega_{TW}=\lambda_0^2-q_0^2/2-sq_0$
 of an {\it extended} wave at
the same wave number. Whether this difference is relevant for the fact that
 the pulse frequencies in the experiments \cite{StKa93} and in
the numerical simulations \cite{BaLu94} are larger than
those of the extended traveling waves is not clear at this point.

\subsection{Second-Order Evolution Equations: Effect of $C$ on the Growth Rate}
\label{s:ord2}
In earlier work Barten {\it et al.} had pointed out that the experimentally observed
 slowing-down of the pulse could be due to the fact that the concentration field leads 
to a reduction of the local {\it growth rate} of the convective amplitude ahead of the
pulse\cite{BaLu94,BaLu94a,BaLu91,LuBa92}. 
This observation motivated the derivation of the extended
Ginzburg-Landau equations (\ref{e:ecglA},\ref{e:ecglC}) \cite{Ri92}. 
Eq.(\ref{e:defsecgl}) shows, however, that changes in the local growth rate do not affect 
the velocity at leading order. Variations in the growth rate come 
in only at $O(\epsilon)$ (cf. (\ref{e:v1})). 
At that order any change in the wave number $q_1$ becomes relevant as well.
Since $q_1$ is determined only at $O(\epsilon^2)$ this implies that one has to go to
 next order in the perturbation expansion in order to capture this effect.

To simplify the calculation it is now
assumed that the leading-order, dispersive effect on the wave number vanishes, 
i.e. $\sigma_j=0$, $j=-1,..2$. In addition, since $h_1$ is small for 
 free-slip-permeable boundary conditions it 
is neglected as well (thus, $\rho_{-1}=0$). Once the 
solvability conditions
at $O(\epsilon)$ are met, the amplitude $A_1$ is determined using the general solution
(\ref{e:defA1gen}). Inserting it into the solvability conditions one then obtains
at $O(\epsilon^2)$
\bea
\frac{d}{dT} \lambda_1 &= &\left(\frac{16}{3}p_r \lambda_0^4 + 2 (2c_r-d_r)\lambda_0^2 
+4 \Lambda S \rho_0 \lambda_0 + 2 (a_r-d_r q_0^2) \right) \lambda_1 \nonumber \\
& &-4d_r q_0 \lambda_0 q_1 +\frac{2176}{945} \lambda_0^7 p_r p_i +\nonumber \\
& & \frac{32}{15}\lambda_0^5 \left( (\rho_2 q+0+c_i)p_r+\frac{2}{3} (c_r-d_r)p_i\right) +\frac{4}{3} \lambda_0^3 \left( (c_r-d_r)(q_0\rho_2+c_i)\right),\\
\frac{d}{dT} q_1 &=& -\frac{4}{3} \lambda_0^2q_1d_r-\frac{8}{3} d_r \lambda_1 \lambda_0 q_0 + \frac{32}{63} p_r (\rho_1 +\frac{1}{5} \rho_2) \lambda_0^6 \nonumber \\
& &+ \left( \frac{4}{15} (\frac{8}{3} c_r -d_r) \rho_1 + \frac{4}{15} (\frac{1}{3} c_r-d_r) \rho_2 -
\frac{128}{45} q_0 p_i d_r + \frac{136}{225} \rho_0 p_r \right) \lambda_0^4 \nonumber \\
& &\frac{4}{3} \lambda_0^3 \rho_0 \rho_1 \Lambda S + 
\frac{2}{3}\left( \frac{1}{3} (4 c_r + d_r) \rho_0 + 2\, (a_r-d_r q_0^2)
-4\,( d_r c_i q_0 + \rho_2 d_r q_0^2) \right) \lambda_0^2 \nonumber \\
& &+2 \lambda_0 \Lambda S \rho_0^2 + 2 (a_r - d_r q_0^2) \rho_0.
\eea
Similar to eqs.(\ref{e:v1},\ref{e:om1}) the equations for $v_2$ and $\omega_2$ 
contain $\lambda_2$ and $q_2$ which are only determined at $O(\epsilon^3)$.

For the equations describing the effect of the concentration mode the coefficients
$\rho_j$ have to be inserted. Eq.(\ref{e:q0}) shows that without dispersion $q_0=0$ and
as a consequence $V_0=s$. For $s=O(1)$ the equation determining $v_1$ will be
linear in $v_1$ (cf. (\ref{e:v1a})) and will therefore not be able to capture 
the hysteretic behavior found in previous numerical simulations \cite{Ri92b}. I therefore
take $s=\epsilon s_1$. In order for the concentration field to remain a small perturbation
one then has to take $h_2=O(\epsilon)$ as well. 
 In addition, for the expansion (\ref{e:defC})
to be valid $\alpha$ and $\delta$ have to be small as compared to $V=\epsilon V_1$. With
these changes the
solvability conditions at $O(\epsilon)$ yield again (\ref{e:la0})-(\ref{e:om1}) with
$V_1$ replaced by $V_1-s_1$ and $V_0$ replaced by $V_1$.  At $O(\epsilon^2)$ one obtains 
\bea
{\frac {d}{dT}}\lambda_1&=&\left ( 4f_rh_2\left(-{\frac {\lambda_0^{2}}{V_{{1}}}}+\frac {\Lambda\,S\alpha\,\lambda_0 }{V_{{1}}^{2}}\right)+
{\frac {16\,p_{{r}}\lambda_0^{4}}{3}}+2\left (2\,c_{{r}}-d_{{r}}\right )\lambda_0^{2}+
2\,(a_{{r}}-d_{{r}}q_0 ^{2})\right )\lambda_1 \nonumber \\
& &-4\,d_{{r}}q_0 q_1 \lambda_0 +{\frac {2176\,\lambda_0^{7}p_{{r}}p_{{i}}}{945}}+
\frac{64}{15}(c_r-d_r) p_{{i}}\lambda_0^{5}\nonumber \\
& &-{\frac {64\,f_{{r}}p_{{i}}h_{{2}}\lambda_0 ^{5}}{45
\,V_{{1}}}}+\frac {4\,f_{{r}}V_{{2}}h_{{2}}\lambda_0 ^{3}}{3\,
V_{{1}}^{2}}+O(V_{{1}}^{-3}), \label{e:loesla1}\\
{\frac {d}{dT}}q_1&=&-\frac{4}{3}d_{{r}}\lambda_0^{2}q_1-
\frac{8}{3}d_{{r}}\lambda_1\lambda_0 
q_0-\frac{128}{45}q_0p_{{i}}d_{{r}}\lambda_0^{4}\nonumber \\
& &+\left ( \frac{2}{9}\left (4\,c_{{r}}+d_{{r}}\right )\lambda_0^{2}
+2\,(a_{{r}}-d_{{r}}q_0^{2})+
{\frac {136\,p_{{r}}\lambda_0 ^{4}}{225}}\right )
\frac{h_{{2}}f_{{r}}\alpha}{V_1^2}
\nonumber \\
& &+\left(\frac{4}{3}\left(a_r-d_{{r}}q_0^{2}\right)\lambda_0^{2}+
{\frac {32\,p_{{r}}\lambda_0 ^{6}}{63}}+
\frac{4}{15}\left (\frac{8}{3}c_{{r}}-d_{{r}}\right)\lambda_0^{4}
\right)\frac{h_{{2}}f_{{r}}\delta}{V_1^2} +O(V_{{1}}^{-3}). \label{e:loesq1}
\eea
In the steady case, one obtains from (\ref{e:la0},\ref{e:q0}) $q_0=0$ and
\be
a_{{r}}=-{\frac {8\,p_{{r}}\lambda_0 ^{4}}{15}}+
\frac{1}{3}\left (d_{{r}}-2\,c_{{r}}+{\frac {2\,f_{{r}}h_{{2}}}{V_{{1}}}}\right )
\lambda_0 ^{2}-{\frac {f_{{r}}\Lambda\,S\alpha\,h
_{{2}}\lambda_0 }{V_{{1}}^{2}}}+O(V_{{1}}^{-3}). \label{e:defars}
\ee
This yields for $\lambda_1$ and $q_1$ 
\bea
\lambda_1& =&-{\frac {16}{63}\frac{\,p_{{i}}\lambda_0 ^{3}\left (34\,
p_{{r}}\lambda_0 ^{2}-21\,d_{{r}}+21\,c_{{r}}\right )}{16\,p_
{{r}}\lambda_0 ^{2}-5\,d_{{r}}+10\,c_{{r}}}}-\nonumber \\
& &\frac{16}{63}\frac{f_{{r}}p_{{i}}h_{{2}}\lambda_0^{3}\left 
(4\,p_{{r}}\lambda_0^{2}-105\,d_{{r}}\right )}
{\left (16\,p_{{r}}\lambda_0^{2}-5\,d_{{r}}+
10\,c_{{r}}\right )^{2}}\frac{1}{V_{{1}}}+O(V_{{1}}^{-2}), \label{e:defla1} \\
q_1&=&-\frac{f_{{r}}h_{{2}}}{d_rV_1^2}\left (\frac{26}{75}\,p_{{r}}\lambda_0^{2}
+\frac{1}{3}(c_r-2d_r)\right )\alpha-\nonumber \\
& &\frac{f_{{r}}h_{{2}}}{105 d_rV_1^2}2\,\lambda_0^{2}\left (8\,p_{{r}}\lambda_0^{2}+7(c_r-d_{{r}})\right )\delta
+O(V_{{1}}^{-3}). \label{e:defq1}
\eea
To determine the velocity $V_1$ the expression for $q_1$ is inserted
into (\ref{e:v1}) with $h_1=0=v_1$. Using (\ref{e:defars}) to replace $\lambda_0^4$
one obtains
\bea
s_{{1}}=V_{{1}}+\left (\left ({\frac {26\,p_{{r}}\lambda_0 ^{2}}{75\,d_{{r}}}}+\frac{1}{3}\left(1+
{\frac {c_{{r}}}{d_{{r}}}}\right)\right )\alpha+\left ({\frac {2}{35}}\left (11-
{\frac {c_{{r}}}{d_{{r}}}}\right )\lambda_0 ^{2}-{\frac {2\,a_{{r}}}{7\,d_{{r}}}}\right )
\delta\right )\frac{h_{{2}}f_{{r}}}{V_{{1}}^2}\nonumber \\
+O(V_{{1}}^{-3}).
 \label{e:defs1}
\eea
Since the amplitude $\lambda_0$ appears only in the terms of order $O(V_1^{-2})$
it can be replaced by the solution of (\ref{e:la0}) in the limit $V_1 \rightarrow \infty$.
In that limit (\ref{e:la0}) describes an unperturbed backward pitch-fork bifurcation.
Of particular interest is the amplitude at the saddle-node bifurcation
and at the Hopf bifurcation
($a_r=0$),
\be
\lambda_{0,SN}^{2}={\frac {5 (d_{{r}}-2\,c_{{r}})}{16\,p_{{r}}}}+
\frac {5\,f_{{r}}h_{{2}}}{8\,p_{{r}}V_{{1}}}+O(V_{{1}}^{-3}), 
\qquad \lambda_{0,H}^{2}=2\lambda_{0,SN}^2.\label{e:la0sn}
\ee
At these special points the velocity is related to the group velocity $s_1$ by
\bea
s_{1}&=&V_{{1,SN}}+\left({\frac {\left (4\,c_{{r}}+61\,d_{{r}}
\right )}{336\,(-p_{{r}})}}\left (2\,c_{{r}}-d_{{r}}\right )\delta+
{\frac {\left (14\,c_{{r}}+53\,d_{{r}}
\right )}{120}}\alpha\right)\frac{h_2f_r}{d_rV_{{1,SN}}^{2}}
+O(V_{{1,SN}}^{-3}), \label{e:defsSN}\\
s_{1}&=&V_{{1,H}}-\left({\frac {\left (2\,c_{{r}}-d_{{r}}\right )
}{28\,(-p_{{r}})}}\left (c_r-11\,d_{{r}}\right)\delta+
{\frac {\left (2\,c_{{r}}-11\,d_{{r}}\right)\alpha}{20}}\right)
\frac{h_2f_r}{d_rV_{{1,H}}^{2}}+O(V_{{1,H}}^{-3}). \label{e:defsH}
\eea
To interpret (\ref{e:defs1},\ref{e:defsSN},\ref{e:defsH}) note that $p_r <0$ and 
$2c_r -d_r >0$ for the pulse to
exist stably in the absence of the concentration mode, $f=0$ (cf. (\ref{e:defars})). 
Therefore,  if $f_r h_2 >0$ the concentration mode slows 
the pulse down when it is first created in the saddle-node bifurcation. This condition, which is satisfied
by the coefficients in the free-slip-permeable case \cite{Ri92a},  
corresponds to the concentration field being advected in such a way that the contribution
to the buoyancy from the concentration is reduced ahead of the pulse. Eq.(\ref{e:defsSN})
 supports therefore the physical interpretation for the slow-down given previously 
\cite{BaLu94,BaLu94a,BaLu91,LuBa92,Ri92}. 

With increasing $a_r$ the situation can, however, change. 
Eq.(\ref{e:defs1}) shows that the contribution to the slow-down 
from the damping term proportional to $\alpha$ decreases with increasing amplitude
 $\lambda_0^2$. The contribution from the diffusive term proportional to $\delta$
can have either sign;
while the term proportional to $a_r$ always leads  to an increase in the velocity with
increasing $a_r$, the effect of the  term proportional to $\lambda_0^2$ depends on the  
ratio $d_r/c_r$.  For 
\be
c_r>11d_r \label{e:crdr}
\ee
its contribution  increases the velocity,
whereas in the opposite case the `trapping' is enhanced with increasing amplitude.
In the immediate vicinity of the saddle-node bifurcation the amplitude $\lambda_0^2$ 
increases more rapidly than $a_r$ and will dominate the behavior. Thus the outcome depends
on the relative strength of damping and diffusion; the pulse velocity will increase
if 
\be
\frac{\alpha}{\delta} > \frac{15}{91} \frac{d_r}{(-p_r)} \left( 11 - \frac{c_r}{d_r} \right)
\label{e:speedup}
\ee
and decrease otherwise. Thus, if $c_r>11d_r$ the
pulse velocity increases with $a_r$ 
independent of the strength of the damping term and, in fact, for sufficiently large $a_r$
the concentration mode even leads to an over-all {\it acceleration} of the pulse  
 (cf. (\ref{e:defsH}) at $a_r=0$), i.e. the pulse is even
faster than the linear group velocity. If (\ref{e:speedup}) is satisfied but not
(\ref{e:crdr}), i.e. $c_r<11d_r$, then the velocity still increases with $a_r$ but remains
below the group velocity as long as $a_r <0$.

These results show that the response of the pulse to the concentration mode is more complex
than suggested by the original picture in which the concentration mode only led to a 
`barrier' of reduced buoyancy  which can `trap' the pulse \cite{BaLu91,Ri92,BaLu94a}.
Since the velocity of the solitary pulse is closely connected to its wave number $q$
(cf. (\ref{e:defv0})) any effect of the concentration mode on the local wave number
 becomes just as 
important. In fact, the dependence of the local frequency on the concentration 
mode, which changes the local wave number, selects a velocity of $O(1)$ (cf.
(\ref{e:defsecgl})). But even when this dispersive effect is absent ($f_i=0$) the variation
of the concentration mode can select a non-zero wave number \cite{Pipriv} somewhat similar
to the selection of the wave number by ramps in the control parameter \cite{KrBe82,PaRi91}.
Eq.(\ref{e:v1}) shows that in the absence of any change in wave number 
the barrier generated by
the advected concentration mode, which is represented by the term proportional to 
$h_2$, would always decrease the pulse velocity.

\section{Numerical Simulations}
\label{s:num}
The perturbation analysis in sec.\ref{s:ord1} and sec.\ref{s:ord2} revealed two distinct mechanisms 
that can lead to
a slow-down of the pulse due to the coupling to the concentration mode. A dispersive 
mechanism which affects the velocity through the wave number and a mechanism that
is based on a suppression of convection ahead of the pulse. The analysis suggests 
that there is a range of parameters over which the pulse is slow. However, due to the 
expansion employed for the calculation of the concentration mode the analysis is strictly 
valid only for 
large velocities and can only be suggestive with regard to the behavior at small velocities. 
To go beyond this limitation the extended Ginzburg-Landau equations (\ref{e:ecglA},\ref{e:ecglC})
are solved numerically. 

\begin{figure}[htb]
\begin{picture}(420,240)(0,0)
\put(0,-45) {\includegraphics{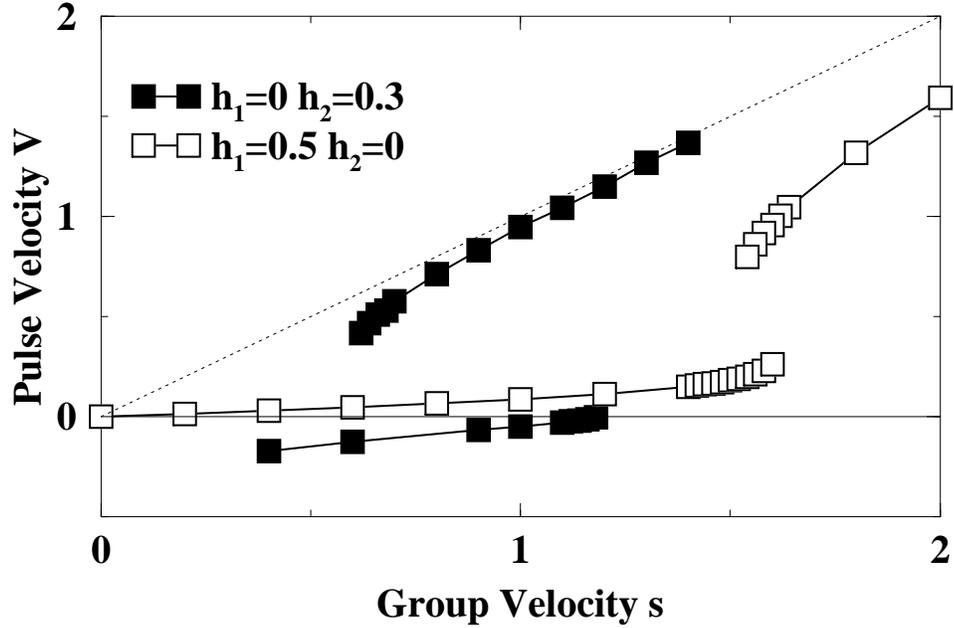}}
\end{picture}
\caption{Numerically determined pulse velocity $V$  as a 
function of the group velocity $s$ for $d_r=0.15+i$, $a=-0.24$, $f=0.5i$,
$c=2.4+2i$, $p=-1.65+2i$, $\delta=0.1$, $\alpha=0.08$ and the indicated values of
$h_{1}$ and $h_2$.
\protect{\label{f:vsfi}}
}
\end{figure}

To study the dispersive `trapping', eqs.(\ref{e:ecglA},\ref{e:ecglC}) are solved with 
$f$ purely imaginary.
The values of the coefficients are chosen as $d_r=0.15+i$, $a=-0.24$, $f=0.5i$,
$c=2.4+2i$, $p=-1.65+2i$, $\delta=0.1$, $\alpha=0.08$
and the   
boundary conditions are periodic. Fig.\ref{f:vsfi} shows the velocity of the pulse
 for $h_1=0$, $h_2 =0.3$ and for $h_1=0.5$, $h_2=0$. As expected from (\ref{e:defsecgl})
 the pulse is slowed down by the concentration mode (the dotted line gives
the velocity without concentration mode, $V=s$). The simulation bears out
even the hysteretic transition to a `trapped' pulse which is suggested by (\ref{e:defsecgl}) 
but outside its range of validity. This transition can be understood intuitively by noting
that the `amount of $C$' generated by the pulse at any given location depends on its velocity.
If the pulse is slow $C(x)$ becomes large and slows it down even more. If the pulse
is fast, however, $C(x)$ is small and the effect on the velocity is small. 

\begin{figure}[htb]
\begin{picture}(420,240)(0,0)
\put(0,-45) {\includegraphics{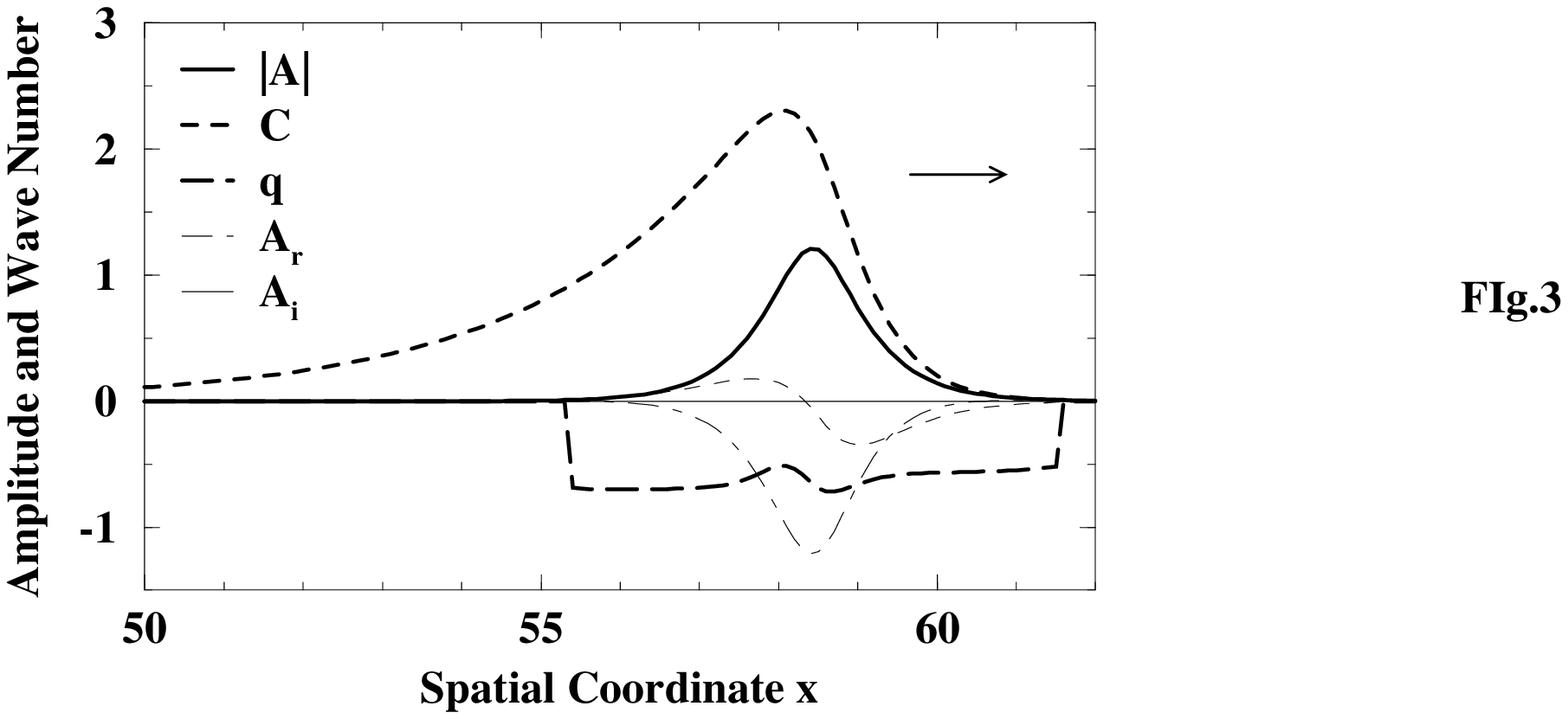}}
\end{picture}
\caption{Typical pulse solution for $f_i \ne 0$ and $h_1\ne 0$. 
The dependence of the frequency on the spatially varying asymmetric concentration 
leads to a shift in the wave number (cf. (\protect{\ref{e:q0}})) which 
changes the velocity of the pulse (cf. (\protect{\ref{e:defv0}})).
\protect{\label{f:puls}}
}
\end{figure}

The terms $h_1 |A|^2$ and $h_2 \partial_x |A|^2$ differ in their symmetry and their
physical interpretation. Whereas the $h_2$-term corresponds to the advection of the 
concentration mode by the wave, the $h_1$-term represents the generation of $C$. 
For the concentration mode to affect the (over-all) wave number it must have an
asymmetric contribution. For $v=0$ the $h_1$-term leads to a symmetric $C$-profile.
Consequently the wave number is increased on one side of the pulse, but decreased by the 
same amount on the other side. The over-all wave number and the pulse velocity
remain unchanged. For non-vanishing velocity, however, the concentration field lags
behind and changes the wave number as illustrated in fig.\ref{f:puls}. 
Due to its symmetry this term can induce `trapping' of the pulse independent of 
its direction of propagation; for $h_2=0$ (\ref{e:defsecgl}) is invariant under
$s \rightarrow -s$ and $V_0 \rightarrow -V_0$. 

By contrast, the $h_2$-term breaks the reflection symmetry at any velocity. For $f_i h_2 >0$, 
the concentration field pushes the pulse to the left independent of its 
direction of propagation. Thus, for positive group velocity $s$ the pulses are
 slowed down, whereas for negative $s$ they are accelerated.  

\begin{figure}[htb]
\begin{picture}(420,240)(0,0)
\put(0,-45) {\includegraphics{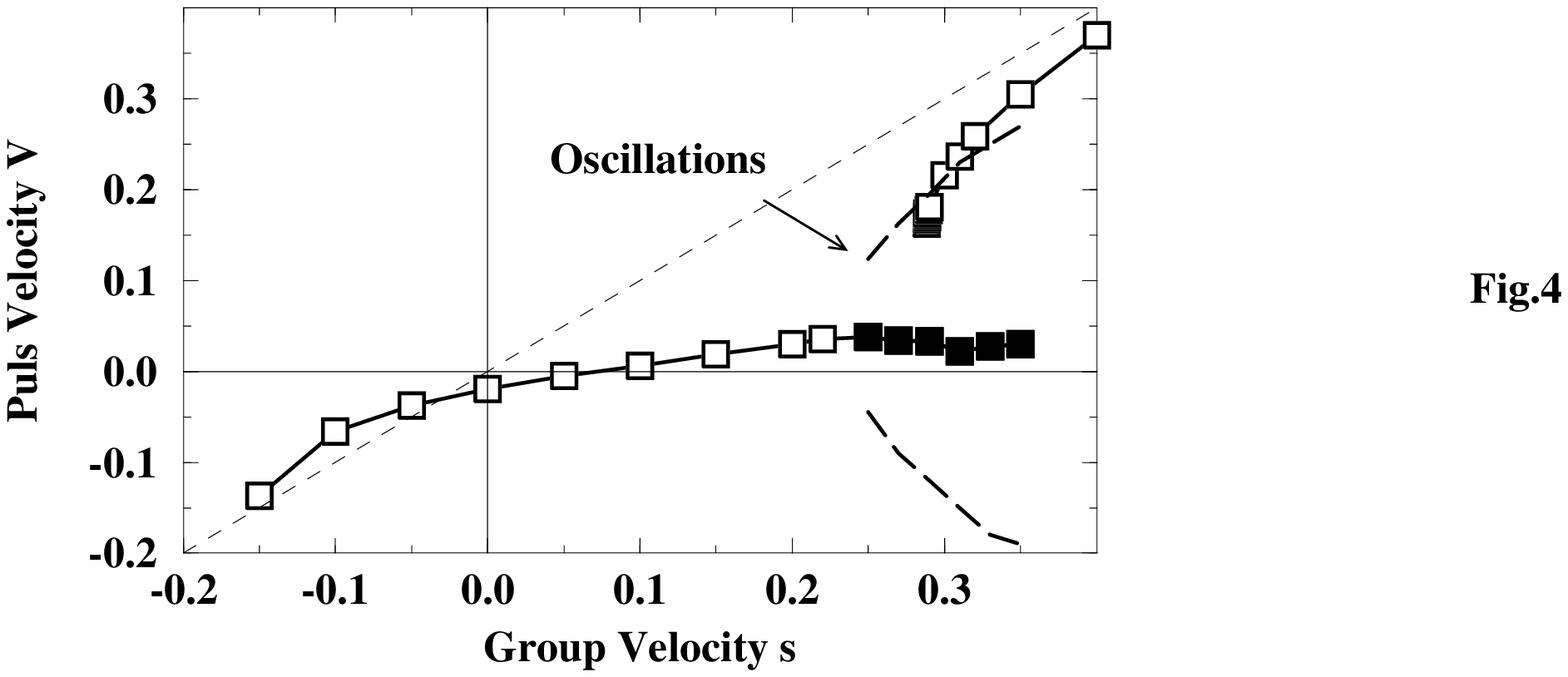}}
\end{picture}
\caption{Numerically determined pulse velocity $V$  as a 
function of the group velocity $s$ for  $a=-0.24$, $d=0.15+i$, $c=2.4+2i$, $p=-1.65+2i$, $\delta=0.1$, $\alpha=0.02$, $f=0.5$, $h_2=0.3$. The dashed lines indicate the minimal
and maximal values of $V$ in the oscillatory regime (solid squares).
\protect{\label{f:vsfr}}
}
\end{figure}

\begin{figure}[htb]
\begin{picture}(420,240)(0,0)
\put(0,-45) {\includegraphics{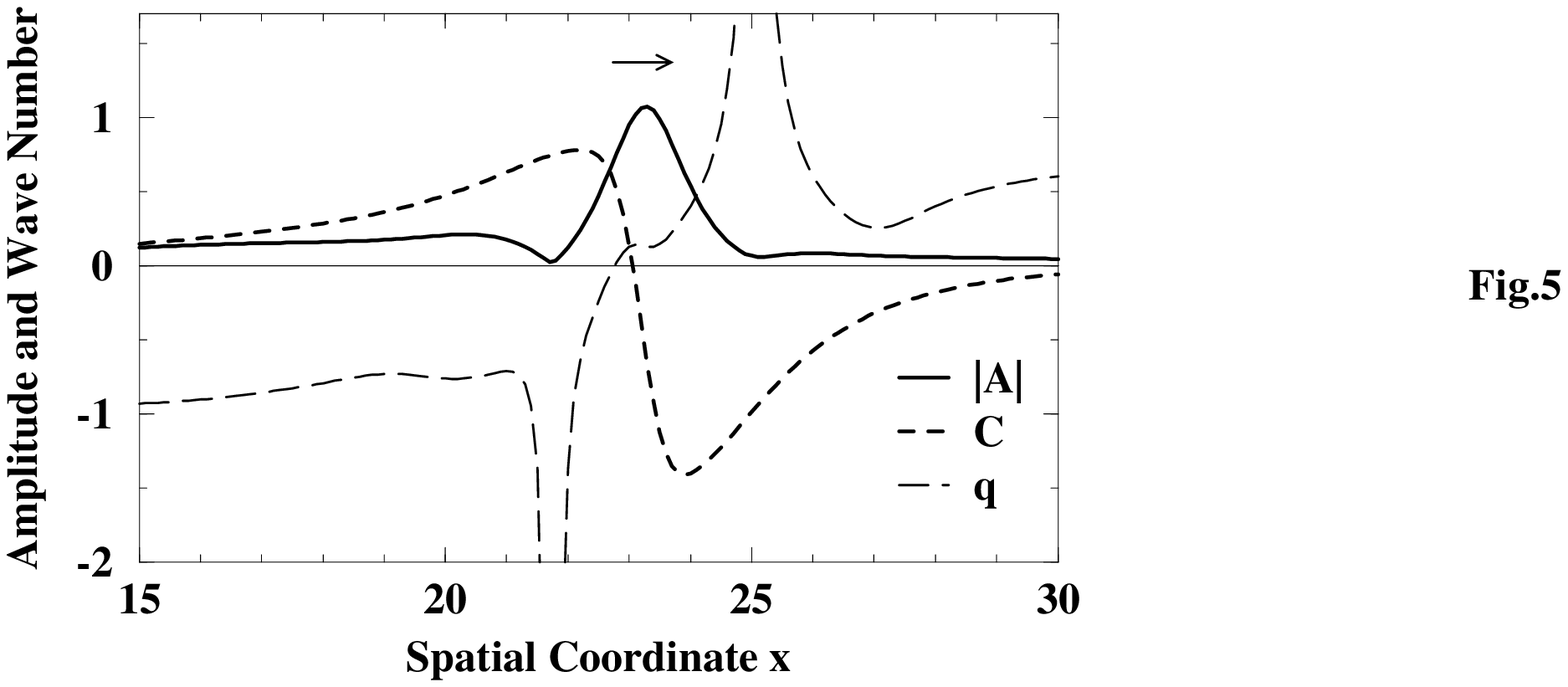}}
\end{picture}
\caption{Pulse solution for $s=0.35$ (other parameters as in fig.\protect{\ref{f:vsfr}}).
The depression in $|A|$ and the diverging wave number indicate 
that a phase slip has just occurred. These phase slips lead to an oscillatory behavior
of the velocity.
\protect{\label{f:pfrh2}}
}
\end{figure}

If the dispersive effects are small the pulse velocity is affected at next order through
the change in the local growth rate given by $f_r CA$. This is the mechanism 
which was pointed out by Barten {\it et al.}
\cite{BaLu94,BaLu94a,BaLu91,LuBa92}. Some numerical simulations of the extended Ginzburg-Landau equations
(\ref{e:ecglA},\ref{e:ecglC}) for this case have been presented previously \cite{Ri92,Ri92b}.
As suggested by (\ref{e:defs1}) the qualitative dependence of the pulse velocity
on the group velocity is similar to that due to dispersion. The results of simulations
for $a=-0.24$, $d=0.15+i$, $c=2.4+2i$, $p=-1.65+2i$, $\delta=0.1$, $\alpha=0.02$,
$f=0.5$, $h_2=0.3$ are shown in fig.\ref{f:vsfr}. Again, for sufficiently large group velocity
a jump-transition from fast to slow pulses is observed. 
In the vicinity of the jump the trapped pulse becomes
unstable to oscillations. It becomes more extended - in particular at its trailing end,
where the
concentration mode leads to an increase in the local growth rate - and phase slips
arise. This is
shown in fig.\ref{f:pfrh2}. Here a pulse is depicted at the time of the phase-slip
process which is marked by the localized amplitude depression and the diverging local wave
number. The phase slips 
lead to an oscillatory behavior of the velocity, the extrema of which are  
indicated by the dashed lines in fig.\ref{f:vsfr}. When the group velocity is increased
further the trailing end develops larger peaks which then grow by themselves and eventually
convection spreads throughout the system. 

Experimentally, the only parameter that can be varied easily is the Rayleigh number, which
enters the growth rate $a_r$. An interesting question is therefore whether the
extended Ginzburg-Landau equations allow a hysteretic transition to fast pulses by
changing $a_r$. Eq.(\ref{e:defs1}) suggests this possibility, but is not valid for
small velocities. This question is investigated for three cases. 

\begin{figure}[htb]
\begin{picture}(420,240)(0,0)
\put(0,-45) {\includegraphics{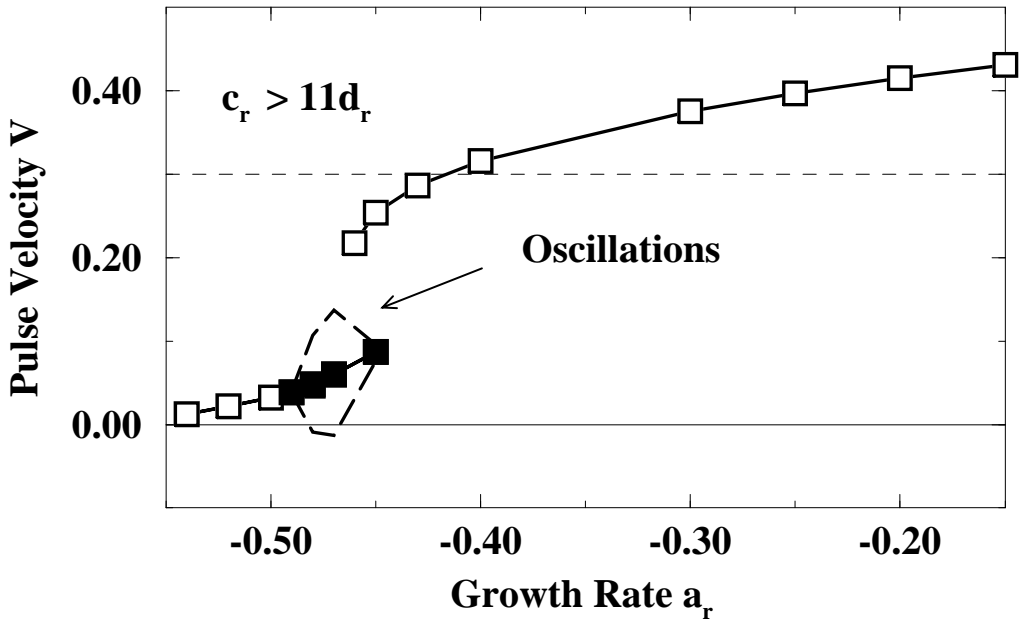}}
\end{picture}
\caption{Numerically determined pulse velocity $V$  as a 
function of the growth rate $a_r$ for $s=0.3$, $d=0.05+0.5i$, $f=0.25$, 
$c=2.4+i$, $p=-1.65$, $\alpha=0.03$, $\delta=0.1$, $h_2=0.3$.
The dashed lines indicate the minimal
and maximal values of $V$ in the oscillatory regime (solid squares).
\protect{\label{f:var1}}
}
\end{figure}

The growth rate
enters (\ref{e:defs1}) in three places. The explicit dependence leads to an increase
in velocity for all $f_r h_2>0$. So does the contribution (via the amplitude $\lambda_0^2$)
from the damping term proportional to $\alpha$. The diffusive term proportional to
$\delta$ can, however, lead to a further decrease in the velocity with increasing 
amplitude. If inequalities (\ref{e:crdr},\ref{e:speedup}) are satisfied, the pulse velocity
is predicted to increase in the immediate vicinity of the saddle-node bifurcation
and reach a value larger than $s$ for $a_r=0$. 
The result of a numerical simulation of this case is presented in 
fig.\ref{f:var1} for $s=0.3$, $d=0.05+0.5i$, $f=0.25$, c=$2.4+i$, $p=-1.65$, $\alpha=0.03$,
$\delta=0.1$, $h_2=0.3$. It shows indeed a hysteretic transition from a trapped 
pulse to an accelerated pulse. Interestingly, the trapped pulse becomes unstable to
oscillations in the vicinity of the transition. Here it is the variation
 in the peak height
of the concentration mode which leads to the oscillations in the velocity. 
Their minimal and maximal values are denoted by dotted lines.

\begin{figure}[htb]
\begin{picture}(420,240)(0,0)
\put(0,-45) {\includegraphics{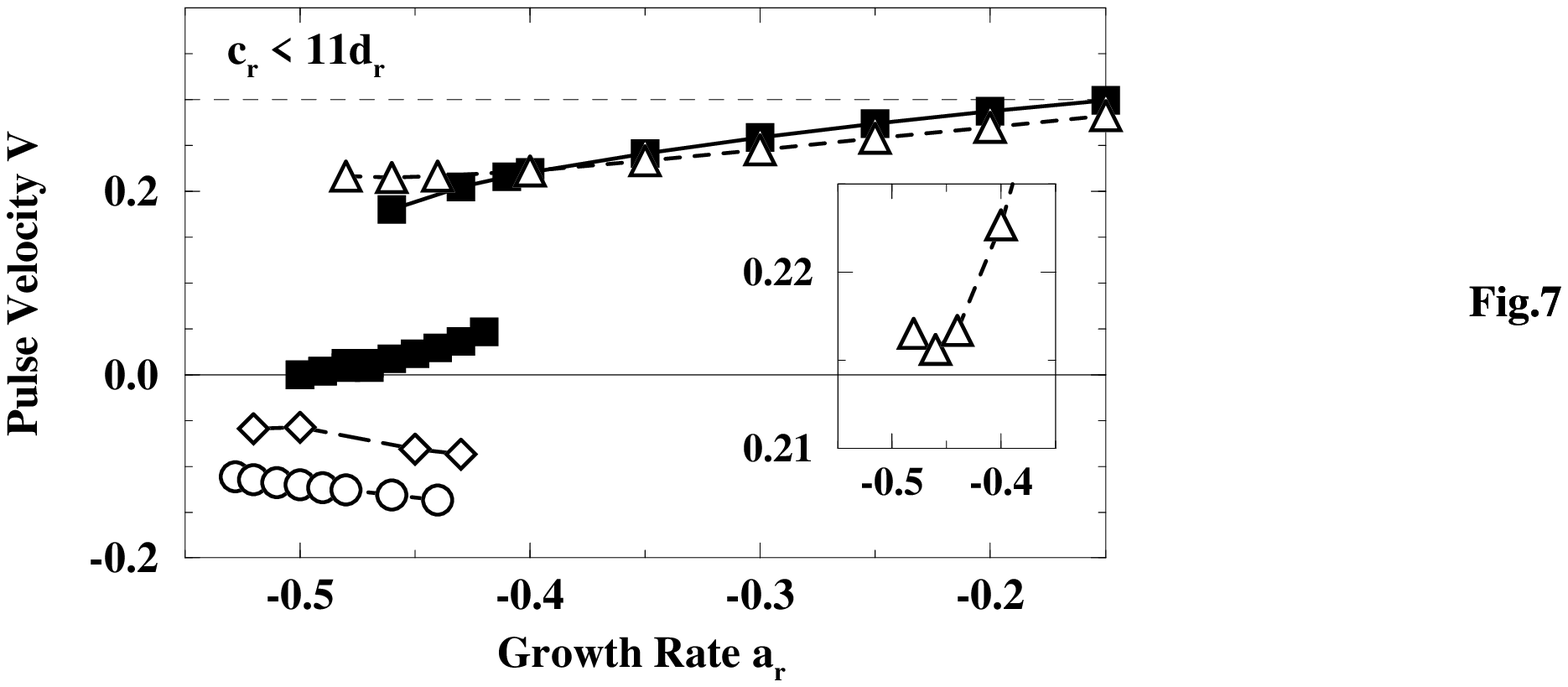}}
\end{picture}
\caption{Numerically determined pulse velocity $V$  as a 
function of the growth rate $a_r$ for various values of $d_r$ and $f_r$: $d_r=0.3$, 
$f_r=0.25$ 
(solid squares),  $d_r=0.6$, $f_r=0.2$ (open triangles), $d_r=0.6$, $f_r=0.25$ (open 
diamonds), $d_r=0.6$, $f_r=0.35$ (open circles). 
Other parameters as in fig.\protect{\ref{f:var1}}. Inset shows initial decrease in velocity
for $d_r=0.6$, $f_r=0.2$ as expected from (\protect{\ref{e:crdr}}).
\protect{\label{f:var2}}
}
\end{figure}

Fig.\ref{f:var2} shows the result of simulations for a case in which $c_r<11d_r$ but
inequality (\ref{e:speedup}) is still satisfied (solid squares, 
$d_r=0.3$, other parameters as 
in fig.\ref{f:var1}). As expected from the perturbation result the pulse
velocity increases in the vicinity of the saddle-node bifurcation ($a_r\approx-0.5$), 
but it remains
below the group velocity up to $a_r=-0.15$. For larger values of $a_r$ the pulse becomes
unstable. Finally, if both inequalities (\ref{e:crdr},\ref{e:speedup}) are violated
the perturbation analysis suggests that the pulse is slowed down even further with increasing
$a_r$. Two such cases are shown in fig.\ref{f:var2} ($d_r=0.6$). 
 For $f_r=0.35$ (open circles)  and for $f_r=0.25$ (open diamonds) 
the pulse is `trapped' and its velocity 
{\it decreases} up to $a_r=-0.43$. 
For larger values of $a_r$ the pulse 
starts to grow and eventually fills the whole cell. For weaker coupling to the concentration
mode ($f_r=0.2$, open triangles) the pulse velocity decreases with increasing $a_r$ 
in the immediate vicinity of the saddle-node bifurcation, but for larger $a_r$ 
the velocity starts to increase.

\section{Conclusion}
\label{s:concl}
In this paper I have investigated the influence of a long-wavelength mode on the dynamics of
a traveling-wave pulse in the perturbed nonlinear Schr\"odinger equation. The study was motivated by 
experimental and numerical results on convection in binary liquids. It showed analytically
that such a mode strongly affects the propagation velocity of such a pulse. In particular,
two distinct mechanisms were identified which can lead to a `trapping' of the pulse. As expected
from numerical simulations of the full Navier-Stokes equations \cite{BaLu94} this can 
be due to a reduction of the local growth rate of the traveling-wave mode. This effect arises, however, only at second
order. Already at first order the frequency of the waves is affected 
by the long-wave mode which induces a change in the wave number and connected with it in
the velocity of the pulse. This effect was neither anticipated by the numerical 
simulations nor by
the experiments. For simplified boundary conditions 
both effects are found to reduce the velocity
unless  dispersion is very large. Within the framework presented here hysteretic transitions
can occur between `trapped' pulses and fast pulses when the local growth rate (i.e. the Rayleigh
number) is increased. So far no such transitions have been found in the experiments or 
the numerical simulations of the Navier-Stokes equations.

An increase in the Rayleigh number in the experiments often leads to long pulses which
cannot be regarded as perturbed solitons anymore. Instead they can be considered
as a bound state of two fronts. The dynamics of such fronts is also strongly
affected by the long-wave mode. In particular, within the extended Ginzburg-Landau
equations discussed here it has been found
that their interaction can change from attractive to repulsive when the direction of propagation of the pulse
changes sign \cite{HeRi94,HeRi95,RiRa95}. Thus, backward traveling pulses are stable
and pulses of different sizes can coexist in agreement with experimental results \cite{Ko94}.
Further studies show \cite{RiRa95}
 that the coexistence of stable pulses found in the numerical simulations \cite{BaLu94}
can also be understood within this framework.

No attempt has been made so far to investigate the effect of increasing the Lewis number
(i.e. the mass diffusion) using full numerical simulations of the Navier-Stokes equations. According to the present analysis the pulse velocity should increase drastically when the 
Lewis number is increased.
Such an investigation would provide a stringent test for the relevance
of the extended Ginzburg-Landau equations to binary-mixture convection.

There have been attempts to find stable localized pulses in two-dimensional convection.
To date these pulses have not proved to be stable but decay after long transients 
\cite{LeBo93}.
The origin of this instability is not understood. In addition to the long-wave concentration mode
a long-wave vorticity mode (mean flow) is expected to be relevant. A related interesting problem
is the dynamics of two-dimensional propagating fronts. Since the concentration mode
slows down certain types of fronts \cite{HeRi95} one may expect that 
it could lead to a transverse instability of such fronts similar to the Mullins-Sekerka
instability in solidification \cite{CrHo93}.

The relevance of the equations discussed in this paper is not confined to convection in 
binary liquids. Equations of the same form have also been derived for waves on the interface
between two immiscible liquids flowing in a channel \cite{ReRe93}. It is conceivable that nonlinear gradient terms arising in the description of other wave systems 
are also due to the elimination of a second mode (e.g. \cite{GoKa94,HeNa94}). 
In regimes in which such a mode becomes slow as 
compared to the wave it would have to be described by a separate dynamical equation 
and could have similarly strong effects on the waves as found in the present paper. 

I gratefully acknowledge very helpful discussions with W.L. Kath. I have also profited from
discussions with M. L\"ucke, L. Pismen and P. Kolodner.
 This work was supported by DOE through grant (DE-FG02-92ER14303) and by
an equipment grant from NSF (DMS-9304397).

\bibliography{/home2/hermann/.index/journal}

\newpage
\appendix
\section{Derivation of the Extended Ginzburg-Landau Equations}
\label{s:appa}
In this appendix I rederive the extended Ginzburg-Landau equations (ECGL) under slightly
more general conditions than previously \cite{Ri92a}. In the previous derivation the
Lewis number $L$ and the growth rate $a$ of the convective mode were taken to be of the
same order. In the limit of very small
growth rate, i.e. for $|a| \ll L$, the concentration mode $C$ can be eliminated 
adiabatically 
from the resulting equations to yield the usual complex Ginzburg-Landa 
equation (CGL). However, with the coefficients calculated in \cite{Ri92a} one does not
recover the correct values of the coefficients in the CGL, not even their 
values in the limit of small Lewis number. 
 The generalization presented here
overcomes this problem and agrees for $|a| \ll L$ quantitatively 
with the CGL. At the same time it shows that the terms $|A|^2$ and
$i\left(A^*\partial_xA-A\partial_xA^*\right)$, 
which in the previous
derivation were found to have vanishing coefficients at quadratic and third order, respectively, become relevant at higher order. 

A simple analysis of (\ref{e:caa},\ref{e:cac}) shows that the coefficients derived in 
\cite{Ri92a} do not lead to the correct coefficients of the single CGL if the concentration
mode is  eliminated adiabatically for $|a| \ll L \propto \alpha$. 
This is due to the fact that any
$L$-dependence of the nonlinear coefficients is shifted to higher-order terms due to
the initial assumption $L=O(\epsilon^2)=O(a)$. Since in the adiabatic elimination the
contributions from the nonlinear terms are divided by $L$, higher-order
terms of the form $L|A|^2$
 become relevant. The goal is to
obtain certain of these contributions without performing the full expansion to
the corresponding high order. 

As pointed out in sec.\ref{s:ecgl}, 
(\ref{e:caa},\ref{e:cac}) are 
reconstituted equations\cite{Ro85}; i.e. the solvability conditions arising at
$O(\epsilon)$, $O(\epsilon^2)$ and $O(\epsilon^3)$ are not satisfied 
separately but are combined into
a single solvability condition. One could imagine such a reconstitution involving
all orders in $\epsilon$. The higher-order terms would then have the form
$L^nA^mC^p$ and suitable spatial derivatives thereof. They could be summed up
into a term $A^mC^p$ (and the respective spatial derivatives) 
with an $L$-dependent coefficient.
Alternatively, one could obtain this
$L$-dependence directly by avoiding the expansion in $L$, thus providing
a summation of certain contributions in $L$ to all orders. This is the approach taken here.
A more detailed analysis is then required to identify
for which of the coefficients the procedure restores the complete $L$-dependence. 

Within the Boussinesq-approximation the basic equations for two-dimensional
convection in a binary mixture at infinite Prandtl number read \cite{ScZi89,ScZi93}
\begin{equation}\left(\begin{array}{ccc}
-\Delta^2&-(1+\psi)\partial_x&-\psi\partial_x\\
-{\cal R}\partial_x&\partial_t-\Delta&0\\
-{\cal R}\partial_x&\partial_t&\partial_t- L\Delta\end{array}
\right)\left(\begin{array}{c}
\phi\\
\theta\\
\eta\end{array}
\right)=\left(\begin{array}{c}
0\\
(\partial_z\phi\partial_x-\partial_x\phi\partial_z)\theta\\
(\partial_z\phi\partial_x-\partial_x\phi\partial_z)(\eta +\theta 
)\end{array}
\right)\label{e:ns}\end{equation}
\noindent
where $\phi$ denotes the stream function, $\theta$ the deviation of the
temperature from the conductive state and $\eta=(c/\psi-\theta)$ 
with $c$ being the deviation of the concentration field from the 
conductive state. The separation ratio is denoted by $\psi$ and ${\cal R}$ and $L$
are the Rayleigh number and the Lewis number, respectively.
For free-slip-permeable boundary conditions the fields are
expanded as \cite{Ri92a}
\bea
{ \phi}&=&\epsilon\,{\rm  e}^{{i}\,{q}\,{x} -i\omega_h t}\,
{\rm sin}(\,{ \pi}\,{z}\,)\,{A}  + \epsilon^{2}\{
{\rm sin}(\,2\,{ \pi}\,{z}\,)\,(\,\frac{1}{2}{\it D_{\phi}} + {\it E_{\phi}}\,
{\rm  e}^{{i}\,{q}\,{x} -i\omega_h t} \,) + \nonumber \\
 & &{\rm sin}(\,{ \pi}\,{z}\,)
\left(\,\frac{1}{2}{\it F_{\phi}} + (\,{\it G_1} + {\it G_2} + {\it G_3}\,)\,
{\rm  e}^{{i}\,{q}\,{x} -i\omega_h t} \,\right)
 \mbox{} + {\rm sin}(\,3\,{ \pi}\,{z}\,)\,{\it H_{\phi}}\,{\rm 
 e}^{{i}\,{q}\,{x} -i\omega_h t} \} \nonumber \\
& &+c.c.+ O(\epsilon^3), \label{e:expp}
\\
{ \theta}&=&\epsilon\,{\rm  e}^{{i}\,{q}\,{x} -i\omega_h t}\,
{\rm sin}(\,{ \pi}\,{z}\,)\,{A}\,{\it \zeta_1}  +
\epsilon^{2}\{
{\rm sin}(\,2\,{ \pi}\,{z}\,)\,\left(\,\frac{1}{2}{\it D_{\theta}} + {\it E_{\theta}}\,
{\rm  e}^{{i}\,{q}\,{x} -i\omega_h t}  \,\right) \nonumber \\
 & & \mbox{} + {\rm sin}(\,{ \pi}\,{z}\,)
\left(\frac{1}{2}{\it F_{\theta}} + {\it \zeta_1}\,(\,{\it G_1} - {\it G_2} + {\it G_3}\,)
\,{\rm  e}^{{i}\,{q}\,{x} -i\omega_h t} \,\right)
 \mbox{} + {\rm sin}(\,3\,{ \pi}\,{z}\,)\,{\it H_{\theta}}\,{\rm 
 e}^{{i}\,{q}\,{x} -i\omega_h t} \}\nonumber \\
& &+c.c.+ O(\epsilon^3), \label{e:expt}
\\
\eta &=&  \epsilon\,\left(\,{\rm e}^{{i}\,{q}\,{x} -i\omega_h t}\,{\rm sin}(\,{ \pi}\,{z}\,)\,{A
}\,{\it \zeta_2} + {\it C}\,{\rm sin}(\,2\,{ \pi}\,{z}\,)\,\right) + 
 \epsilon^{2}\{{\rm sin}(\,2\,{ \pi}\,{z}\,)\,
\left(\,\frac{1}{2}{\it  D_{\eta}} + {\it E_{\eta}}\,
{\rm e}^{{i}\,{q}\,{x} -i\omega_h t}\right) + \nonumber \\
 & & \mbox{}{\rm sin}(\,{ \pi}\,{z}\,)
\left(\frac{1}{2}{\it F_{\eta}} + {\it \zeta_2}\,(\,{\it G_1} + {\it G_2} - {\it G_3}\,)
\,{\rm e}^{{i}\,{q}\,{x} -i\omega_h t} \right)
 \mbox{} + {\rm sin}(\,3\,{ \pi}\,{z}\,)\,{\it H_{\eta}}\,
{\rm  e}^{{i}\,{q}\,{x} -i\omega_h t} \}\,) \nonumber \\
& &+ c.c.+ O(\epsilon^3). \label{e:expe}
\eea
The Rayleigh number is expanded as
\be
{\cal R}=R_0(q)+\epsilon R_1+\epsilon^2 R_2 \qquad \mbox{ with }   \qquad
{\it R_0}={\displaystyle \frac { \left( \! \,{q}^{2
} + { \pi}^{2}\, \!  \right) ^{3}\,(\,1 + {L}\,)}{{q}^{2}\,(\,1
 + { \psi}\,)}}.\label{e:defR}
\ee
At this point the expansion wave number $q$ is performed is left undetermined. 
This allows a consistency check of the resulting coefficients since certain
higher-order coefficients have to correspond to derivatives of lower-order
coefficients with respect to the wave number $q$ (see appendix \ref{s:appb}). 
Eventually $q$ will be chosen to be the critical wave number $q_c=\pi/\sqrt{2}$.
The separation ratio $\psi$ and the Hopf frequency are expressed as
\be
\psi = -\frac{\Omega^2}{\Omega^2+1},\qquad
\omega_h=\omega (q^2+\pi^2) \mbox{  with  } \omega = \sqrt{\Omega^2 -L^2 +L \Omega^2}.
\ee
The linear eigenvector $(1,\zeta_1,\zeta_2)^t$ of the convective mode 
is given by
\be
\zeta_1=\frac{i (\Omega^2+1)(1+L)(q^2+\pi^2)^2}{(1-i\omega)q}, \qquad
\zeta_2=\zeta_1\frac{1}{L-i\omega}.
\ee
As mentioned above, the Lewis number $L$ is not expanded. Thus, already at $O(\epsilon)$ a damping term $-4\pi^2 L C$ arises 
formally in the solvability condition for $C$. At this order 
no time-derivative arises which would balance this term. 
Since $L$ is eventually taken to be small
  ($O(\epsilon)$) it is reasonable to defer the balancing until
 all solvability conditions for $C$ are reconstituted into a single
 equation.

At $O(\epsilon^2)$ one
obtains 
\bea
{\it F_{\theta}}&=&0, {\it F_{\phi}}=0, {\it F_{\eta}}=0,\qquad
{\it E_{\theta}}=0, {\it E_{\phi}}=0, {\it E_{\eta}}=0,\label{e:defF}\\
 {\it D_{\theta}}&=&{\displaystyle \frac {1
}{2}}\,{\displaystyle \frac { \left( \! \,{q}^{2} + { \pi}^{2}\,
 \!  \right) ^{2}\, \left( \! \,{ \Omega}^{2} + 1\, \!  \right) 
\,|A|^2}{ \left( \! \,{L} - { \Omega}^{2} - 1\, \! 
 \right) \,{ \pi}}},\qquad
 {\it D_{\phi}}={\displaystyle \frac {1}{16}}\,
{\displaystyle \frac {{ \Omega}^{2}\, \left( \! \,{\frac {{ 
\partial}}{{ \partial}{X}}}\,{\it C}\, \!  \right) }{{ \pi}^{4}
\, \left( \! \,{ \Omega}^{2} + 1\, \!  \right) }}\, \! 
,\\ 
{\it G_2} & =&  \,{\displaystyle \frac {1}{4}}\,{\displaystyle \frac {{q
}^{2}\,(\,iL-{ \omega})\,{A}\,{\it R_1}}{(\,1 + {L}\,)\,
{ \omega}\, \left( \! \,{q}^{2} + { \pi}^{2}\, \!  \right) ^{3}\,
 \left( \! \,{ \Omega}^{2} + 1\, \!  \right) }} - {\displaystyle 
\frac {1}{4}}\,{\displaystyle \frac {{i}\,{ \Omega}^{2}\,{ \pi}\,
{q}^{2}\,{\it C}\,{A}}{{ \omega}\, \left( \! \,{ \Omega}^{2} + 1
\, \!  \right) \, \left( \! \,{q}^{2} + { \pi}^{2}\, \!  \right) 
^{3}}} \nonumber \\
& & \mbox{} - {\displaystyle \frac {1}{2}}\,{\displaystyle 
\frac { \left( \! \, - { \pi}^{2}\,{L} + 2\,{L}\,{q}^{2} - {i}\,{
q}^{2}\,{ \omega}\, \!  \right) \, \left( \! \,{\frac {{ \partial
}}{{ \partial}{X}}}\,{A}\, \!  \right) }{{ \omega}\,{q}\, \left( 
\! \,{q}^{2} + { \pi}^{2}\, \!  \right) }} \label{e:defG2}
,\\
{\it G_3}&=& 
{\displaystyle \frac {1}{4}}\,{\displaystyle \frac {{i}\,{q}^{2}
\,{A}\,{\it R_1}}{{ \omega}\, \left( \! \,{ \Omega}^{2} + 1\, \! 
 \right) \, \left( \! \,{q}^{2} + { \pi}^{2}\, \!  \right) ^{3}}}
 - {\displaystyle \frac {1}{4}}\,{\displaystyle \frac {{ \pi}\,{q
}^{2}\,(\,  { \omega} + {i}\,{L}\,)\,{\it C}\,{A}}{{ \omega}\,
 \left( \! \,{ \Omega}^{2} + 1\, \!  \right) \, \left( \! \,{q}^{
2} + { \pi}^{2}\, \!  \right) ^{3}}} \nonumber \\
& & \mbox{} - {\displaystyle \frac {1}{2}}\,{\displaystyle 
\frac { \left( \! \, - { \pi}^{2} - { \pi}^{2}\,{L} + 2\,{q}^{2}
 + 2\,{L}\,{q}^{2} - 3\,{i}\,{q}^{2}\,{ \omega} + {i}\,{ \omega}
\,{ \pi}^{2}\, \!  \right) \, \left( \! \,{\frac {{ \partial}}{{ 
\partial}{X}}}\,{A}\, \!  \right) }{{ \omega}\,{q}\, \left( \! \,
{q}^{2} + { \pi}^{2}\, \!  \right) }}, \label{e:defG3}
\\
{\it H_{\phi}}&=&  4\,{\displaystyle \frac {{i}\,(\, 19\,{i} + 3\,{ \omega}\,)\,{ 
\Omega}^{2}\,{A}\,{\it C}}{ \left( \! \,{ \Omega}^{2} + 1\, \! 
 \right) \,{ \pi}^{3}\,k_1}},  
\\
{\it H_{\eta}}&=& - \,{\displaystyle \frac {{i}\,\sqrt {2}\,{A}\,
{\it C}\,(\,27\,{L} - 6832 + 1083\,{i}\,{ \omega}\,)}{k_1
}},  
\\
{\it H_{\theta}}&=& - 27\,{\displaystyle \frac {{i}\,(\,1 + {L}\,)\,
\sqrt {2}\,{ \Omega}^{2}\,{A}\,{\it C}}{k_1}} \label{e:defH}
\eea
with
\be
k_1 =  513 \, (L^2-\Omega^2-L \Omega^2 ) + 3249 \, \omega^2 +
20496 \, i \omega (L + 1) - 129808 \, L.
\ee
Note that for simplicity the expressions $H_\phi$, $H_\theta$ and $H_\eta$ have
been given only for $q=q_c$. In the expressions for $G_2$ and $G_3$, however,
the $q$-dependence has been retained for use in appendix \ref{s:appb}. 
The amplitudes $G_1$ and $D_\eta$
can be set to 0 since they only renormalize the amplitudes $A$ and $C$, respectively.
Combining the solvability conditions for $A$ and $C$ arising at $O(\epsilon)$,
$O(\epsilon^2)$ and $O(\epsilon^3)$ one obtains the ECGL
\bea
\partial_tA+(s+s_2C)\partial_xA&=&d\partial_x^2A+(a
+fC+f_2C^2+f_3\partial_xC)A+cA|A|^2,\label{e:caaapp}\\
 \partial_tC&=&\delta \partial_x^2C - \alpha C+ h_1 |A|^2 + 
h_2\partial_x|A|^2+h_3C|A|^2+ \nonumber \\
 & &ih_4(A^*\partial_xA-A\partial_xA^*).\label{e:cacapp}
\eea
The coefficients are given by
\bea
{a}&= & \mbox{}  {\displaystyle \frac {1}{9}}\,{\displaystyle 
\frac {\epsilon\, \left( \! \,  { \omega} + {i}\,{L} - {i}
\,{ \Omega}^{2}\, \!  \right) \,{\it R_2}}{{ \pi}^{2}\,{ \omega}\,
 \left( \! \,{ \Omega}^{2} + 1\, \!  \right) }},\label{e:defa}
\\
{s}&=& \sqrt {2}\,{ \omega}\,{ \pi}, \qquad
{\it s_2}= - \,{\displaystyle \frac {2}{27}}\,{\displaystyle 
\frac {(\,{i}\,{ \omega} - 1\,)\,{ \Omega}^{2}\,\sqrt {2}\,\epsilon
}{ \left( \! \,{ \Omega}^{2} + 1\, \!  \right) \,{ 
\pi}^{2}\,{ \omega}}}, \label{e:defs}
\\
{d}&=&2\,(\,1 + {L}\,)\,\epsilon + {\displaystyle \frac {{i}
\, \left( \! \,{L}^{2} - {L}\,{ \Omega}^{2} + 2\,{L} - { \Omega}
^{2}\, \!  \right) \,\epsilon}{{ \omega}}},
\\
{c}&=& - \,{\displaystyle \frac {1}{24}}\,{\displaystyle 
\frac { \left( \! \,4\,{L}^{3} - {L}^{2}\,{ \Omega}^{2} - 3\,{ 
\Omega}^{4}\,{L} - 3\,{ \Omega}^{4}\, \!  \right) \,\epsilon
\,{ \pi}^{2}}{{ \Omega}^{2}\,{ \omega}^{2}\, \left( \! \,{L} - {
 \Omega}^{2} - 1\, \!  \right) }}
 \mbox{} + {\displaystyle \frac {1}{24}}\,{\displaystyle 
\frac {{i}\, \left( \! \,4\,{L}^{2} + 3\,{L}\,{ \Omega}^{2} - 3\,
{ \Omega}^{4}\, \!  \right) \,{ \pi}^{2}\,\epsilon}{{ 
\Omega}^{2}\,{ \omega}\, \left( \! \,{L} - { \Omega}^{2} - 1\,
 \!  \right) }},
\\
{f}&=&  -\,{\displaystyle \frac {1}{27}}\,{\displaystyle \frac {
 3\,{ \Omega}^{4}+ {L}^{2}\,{ \Omega}^{2} + 3\,{ \Omega}^{4}\,{L} 
-4\,{L}^{3} }{{ \pi}\, \omega^2 \, \left( \! \,{ \Omega}^{2} + 1\,
 \!  \right) }} - {\displaystyle \frac {1}{27}}\,{\displaystyle 
\frac {{i}\, \left( \! \,4\,{L}^{2} + 3\,{ \Omega}^{2}\, \! 
 \right) }{{ \pi}\, \left( \! \,{ \Omega}^{2} + 1\, \!  \right) 
\,{ \omega}}},\label{e:deff}
\\
{\it f_2}&=&  {\displaystyle \frac {1}{972}}\,
{\displaystyle \frac {{i}\,{ \Omega}\, \left( \! \,255\,{ \Omega}
^{2} + 1651\,{i}\,{ \Omega} + 1708\, \!  \right) \,(\,{ \Omega}
 + {i}\,)\,\epsilon}{{ \pi}^{4}\, \left( \! \,{ \Omega}^{2}
 + 1\, \!  \right) ^{2}\,(\, + 427\,{i} + 57\,{ \Omega}\,)}}  
 \nonumber \\
 & & -{\displaystyle \frac {1}{1944}}\,{\displaystyle \frac {{i}\,
\epsilon\,{ \Omega}\, \left( \! \,14535\,{ \Omega}^{4} +
240342\,{i}\,{ \Omega}^{3} - 1119241\,{ \Omega}^{2} - 778848\,{i}
\,{ \Omega} - 1263066\, \!  \right) }{{ \pi}^{4}\, \left( \! \,{ 
\Omega}^{2} + 1\, \!  \right) ^{2}\,(\, 427\,{i} + 57\,{ \Omega
}\,)^{2}}}\,{L} \nonumber \\
 & &  + {\rm O}(\,{L}^{2}\,),
\\
{\it f_3}&=&-{\displaystyle \frac {1}{864}}\,{\displaystyle \frac {
 \left( \! \,32\,{L}^{2} - 32\,{i}\,{ \omega}\,{L} - 32\,{L}\,{ 
\Omega}^{2} + 48\,{ \Omega}^{2} - 21\,{i}\,{ \omega}\,{ \Omega}^{
2}\, \!  \right) \,\sqrt {2}\,\epsilon}{ \left( \! \,{ 
\Omega}^{2} + 1\, \!  \right) \,{ \omega}\,{ \pi}^{2}}},
\\
{ \alpha}&=&4\,{ \pi}^{2}\,\frac{L}{\epsilon}, \qquad
\delta=\epsilon\, \frac {27}{64}
\left( \Omega^2 \, (1+L) +\frac{64}{27} L  \right), \label{e:defdelta}
\\
{\it h_1}&=&
-{\displaystyle \frac {9}{2}}\,{\displaystyle \frac {{ \pi}^{5}\,
 \left( \! \,{ \Omega}^{2} + 1\, \!  \right) \,{L}}{{ \Omega}^{2}
\, \left( { \Omega}^{2} + 1\, \! -\! \,{L}   \right) }},\qquad
{\it h_2}=-{\displaystyle \frac {3}{8}}\,{\displaystyle \frac {\epsilon
\, \left( \! \,{ \Omega}^{2} + 1\, \!  \right) \,{ 
\pi}^{4}\, \left( \! \,5\,{ \Omega}^{2} + 2\, \!  \right) \,
\omega \,\sqrt {2}}{{ \Omega}^{2}\,\left( \! { \Omega}^{2} + 1\, -\,{L}  \!  \right) }},\label{e:defh2}
\\
{\it h_3}&=&{\displaystyle \frac {1}{12}}\,{\displaystyle 
\frac { \left( \! \,{ \Omega}^{2} + 4\, \!  \right) \,{ \pi}^{2}
\,\epsilon}{{ \Omega}^{2} + 1}} - \nonumber \\
& &{\displaystyle \frac {1}{36}}\epsilon\,{ \pi}^{2}
 \frac{ \,574389\,{ \Omega}^{6} + 28147427\,{ \Omega}^{4}
 + 51487378\,{ \Omega}^{2} + 25526060\, } {{ \Omega}^{2}
\left( \! \,{ \Omega}^{2} + 1\, \!  \right) ^{2}\, \left( 
\! \,3249\,{ \Omega}^{2} + 182329\, \!  \right) \, \! \! }
  {L} + {\rm O}(
\,{L}^{2}\,)
,\\
{\it h_4}&=& - \,{\displaystyle \frac {3}{2}}\,{\displaystyle 
\frac {\epsilon\, \left( \! \,{ \Omega}^{2} + 1\, \! 
 \right) \,{ \pi}^{4}\,{L}\,\sqrt {2}}{{ \Omega}^{2}\, \left( \! 
\,{L} - { \Omega}^{2} - 1\, \!  \right) }}.\label{e:defh4}
\eea
In these expressions $q$ has been set equal to $q_c$. Note that $h_1$ and
$h_4$ are then proportional to the Lewis number and are therefore of higher order
in agreement with the previous calculation \cite{Ri92a}.

Without a further discussion it is not clear to which order in $L$ the coefficients
derived above are actually correct; in general the reconstitution does not yield {\it all}
contributions. Consider the expansion with $L$ taken explicitly to be
$O(\epsilon)$, i.e. $L=\epsilon L_1$.
The procedure involves essentially three steps.
First the equations are expanded in a straightforward way. For the relevant modes
$A$ and $C$ this yields evolution equations containing also the other, stable Fourier modes
like $G_i$ etc. In a second step the stable modes are eliminated adiabatically using 
equations like (\ref{e:defF})-(\ref{e:defH})\footnote{I ignore in 
this discussion the possible effect of 
other critical concentration modes.}. This leads to equations of the form
\be
\partial_t A = A\,{\cal F}(|A|^2,C,\partial_x,\partial_t,L_1), \qquad  
\partial_t C= {\cal G}(|A|^2,C,\partial_x,\partial_t,L_1),\label{e:evolAC}
\ee
where the form of the equation is dictated by symmetry. The functions ${\cal F}$ and 
${\cal G}$ stand
for invariant polynomials in their arguments. The notation indicates that 
$A$ and $A^*$ appear only with equal powers as required by translation symmetry. 
The dependence of ${\cal F}$ and ${\cal G}$ on $R_1$ is not indicated.
 Many of the high-order terms differ from the lower-order terms only in their
powers in $L_1$. By not expanding in $L$ all these terms are summed up into
an $L$-dependent coefficient of the lowest-order term with the same structure regarding
the amplitudes $A$ and $C$ and their derivatives. 

In a third step the time-derivatives in the nonlinear terms are
replaced recursively using (\ref{e:evolAC}). If only amplitudes and
derivatives are counted, this introduces terms of lower order 
since the evolution equations have
the form\footnote{Of course, the coefficients $a$ and $\alpha \propto L$ are small; 
therefore the true order of the respective terms remains the same.}
\be 
\partial_t A = a A + ..., \qquad \partial_t C = - \alpha C + ...\label{e:ACrep}
\ee
As a consequence, the replacement of $\partial_t A$ contributes 
to the $R_1$-dependence ($a \propto R_1$)
and the replacement of $\partial_t C$ to the $L$-dependence of the 
corresponding lower-order term.
Even if no expansion in $L$ is performed, the latter contributions are not obtained unless
the initial expansion is taken all the way to the order of the term 
involving the time-derivative. Thus, due to the third step the above procedure
 will not sum up the $L$-dependence to all orders 
unless the coefficient in question has no contribution from higher-order terms containing
time-derivatives $\partial_t C$. 

Since the elimination of $\partial_t C$ in a given
term introduces a factor $C$, which cannot be eliminated by further substitutions,
low-order coefficients of terms containing only factors of $A$ and its derivatives will not
be affected by the replacement of $\partial_t C$ in the higher-order terms and are 
expected to be correct to all orders in $L$. 
Coefficients of terms containing a factor $C$, however,
may be affected and may therefore only be correct to the order of the initial expansion.
The terms omitted in (\ref{e:ACrep}) do not affect this
argument since they do not decrease the order in terms of $A$, $C$ and their derivatives.
For instance, they lead to replacements like $\partial_t A \rightarrow -s\partial_x A$
or $\partial_t C \rightarrow h_1 |A|^2$.

The above arguments suggest that the coefficients
$s$, $c$, $h_1$, $h_2$, and $h_4$ as given in (\ref{e:defa})-(\ref{e:defh4}) are correct
to all orders in $L$. It is therefore expected that the adiabatic elimination of $C$
for $|a| \ll L$ will yield the same value for the cubic coefficient of $A$ as found
in the direct derivation of the CGL.

\section{Elimination of $C$ and Ward Identities}
\label{s:appb}
In this appendix it is demonstrated that the coefficients as calculated in appendix \ref{s:appa}
are consistent with those of the usual CGL in the appropriate limit and satisfy the 
relevant Ward identities\cite{Am78}. 

To show the consistency with the usual CGL, $C$ is adiabatically
eliminated for $\epsilon \ll L$. In the previous expansion \cite{Ri92a} this 
 did not lead to the correct
cubic coefficient in the CGL since for $\epsilon \ll L$ higher-order
terms of the form $L |A|^2$ become relevant as well, which are of higher order in the previous
expansion. They are kept in the expansion
presented in appendix \ref{s:appa}. 
To wit, $h_1$ is of $O(L)$ and vanished therefore in the previous calculation. 
In the present approach it gives an $O(1)$ contribution to $C$,  
\be
{\it C_{ad}}={\displaystyle \frac {9}{8}}\,{\displaystyle \frac {
|A|^2{ \pi}^{3}\, \left( \! \,{ \Omega}^{2} + 1\, \! 
 \right) }{{ \Omega}^{2}\, \left( \! \,{L} - { \Omega}^{2} - 1\,
 \!  \right) }} + O(\partial_x |A|^2).
\ee
Inserting $C_{ad}$ into (\ref{e:caaapp}) gives an additional cubic term in $A$.
The two cubic terms combined yield the purely imaginary cubic coefficient of the
usual Ginzburg-Landau equation
\be
c_{ad}=-\frac{i}{8}\frac{\pi^2}{\omega},
\ee
in agreement with previous work \cite{Kn86}. 

An additional test of the coefficients calculated in
appendix \ref{s:appa} is possible using the Ward identities which relate coefficients
of gradient terms with $q$-derivatives of other, lower-order coefficients \cite{Am78}.
These identities arise due to the fact that the Ginzburg-Landau equations can 
be derived by an expansion around any wave number $q$ on the neutral curve. Of 
course, in an expansion around a wave number different than that corresponding to the 
minimum of the neutral curve the group velocity parameter $s$ is complex and
the expansion of the Rayleigh number contains also the linear term $R_1$ 
(cf. (\ref{e:defR})). Spatially periodic solutions
with wave numbers $q_1 \equiv q + \Delta q$ differing slightly from $q$ 
can then be determined in two ways. 
Either $q$ is changed to $q_1$ or the difference $\Delta q$ in wave number is  
absorbed in the slow wave number $Q$. In the former case the coefficients
in the Ginzburg-Landau equation change. In the latter case the
gradient terms come into play. 
Any physical quantity $P$ like the stream function $\phi$ depends  
on the expansion wave number $q$ as well as the slow wave number $Q$. 
Since both approaches have to lead to the
same final result one obtains the invariance condition
\be
\frac{d}{dQ}P(q-\epsilon Q,Q)=0.
\ee
For definiteness, consider a specific Fourier mode of the stream function $\phi$
(cf.~(\ref{e:expp})),
\be
\phi_{q,\omega_h}= e^{i(q-\epsilon Q)x-i(\omega_h+\epsilon\hat{\Omega}) t} \sin(\pi z) 
\left( \epsilon {\cal A}e^{iQX+i\hat{\Omega} T} +\epsilon^2 (G_1 + G_2 + G_3) + O(\epsilon^3) \right).
\ee
From (\ref{e:defG2},\ref{e:defG3}) one finds
\be
G_i = e^{iQX+i\hat{\Omega} T} g_i, \qquad \mbox{ with } \qquad g_i=const.  
\ee
Since $G_1$ can be set to zero one obtains
\be
\frac{d}{dQ}{\cal A} =- \epsilon \frac{d}{dQ}( g_2 + g_3) = 
\epsilon \rho {\cal A} + O(\epsilon^2) \label{e:invar}
\ee
with
\be
\rho = - \frac{(\pi^2-3q^2)}{(q^2+\pi^2)q}.
\ee
The concentration mode $C$ is independent of $Q$ since it is not the amplitude of a
periodic function.

Conditions for the coefficients in the extended Ginzburg-Landau equations (\ref{e:caaapp},\ref{e:cacapp})
arise from taking the
$Q$-derivative of (\ref{e:caaapp},\ref{e:cacapp}) and inserting the invariance condition (\ref{e:invar}). It is important
to note that the derivative has to be taken at fixed physical parameters 
like the Rayleigh number.
Thus, $R_1$ has to be replaced by $({\cal R}-R_0(q))/\epsilon$. This implies that the
 linear coefficient $a(q)$, which is proportional to $R_1$, satisfies
\be
\frac{d}{dQ}a(q-\epsilon Q) \equiv -\epsilon \frac{d}{dq} a(q-\epsilon Q) = O(1),
\ee
whereas the $Q$-derivative of all other coefficients is $O(\epsilon)$. 
Applied to the equation for the convective amplitude, (\ref{e:caaapp}), 
the invariance condition
leads then to the expected relations
\bea
s = i \epsilon \frac{d}{dq}a, \label{e:wards},\qquad s_2 = i \frac{d}{dq} f.
\eea
The $Q$-derivative of (\ref{e:cacapp}) yields
\be
2 h_4 = \frac{d}{dq} h_1 - (\rho + \rho^*) h_1 + O(\epsilon). \label{e:wardh4}
\ee
At first sight it might be surprising that $h_4$ is not just proportional to the derivative
of $h_1$. However, since the eigenvector $(1,\zeta_1,\zeta_2)^t$, which connects the
amplitude $A$ with the physical quantities, depends explicitly on the wave number $q$
additional terms like $(\rho + \rho^*) h_1$ have to be expected in general.

To check the Ward identities (\ref{e:wards},\ref{e:wardh4}) the 
$q$-dependence of $a$, $f$ and $h_1$ has to be kept. The calculation presented 
in appendix \ref{s:appa} yields
\bea
{\it a(q)}&=&-{\displaystyle \frac {1}{8}}\,{\displaystyle 
\frac {{i}\,{\it R_1}^{2}\,{ \Omega}^{2}\, \left( \! \,{L} - { 
\Omega}^{2} - 1\, \!  \right) \,{q}^{4}\,\epsilon}{{ \omega
}^{3}\, \left( \! \,{ \Omega}^{2} + 1\, \!  \right) ^{2}\,
 \left( \! \,{q}^{2} + { \pi}^{2}\, \!  \right) ^{5}}} + 
{\displaystyle \frac {1}{2}}\,{\displaystyle \frac {{q}^{2}\,
 \left( \! \, { \omega} + {i}\,{L} - {i}\,{ \Omega}^{2}\, \! 
 \right) \,}{{ \omega}\, \left( \! \,{ \Omega}^{2} + 1\,
 \!  \right) \, \left( \! \,{q}^{2} + { \pi}^{2}\, \!  \right) ^{
2}}}\left({\it R_1}+\epsilon R_2\right), \nonumber \\
\label{e:defaq}
\\
f(\,{q}\,)&=&  \,{\displaystyle \frac {1}{2}}\,
{\displaystyle \frac { \left( \! \,2\,{ \pi}^{2}\,{L}^{3} + {q}^{
2}\,{L}^{2}\,{ \Omega}^{2} - { \Omega}^{4}\,{ \pi}^{2}\,{L} - {q}
^{2}\,{ \Omega}^{4}\,{L} - { \Omega}^{2}\,{L}^{2}\,{ \pi}^{2} - {
q}^{2}\,{ \Omega}^{4} - { \pi}^{2}\,{ \Omega}^{4}\, \!  \right) 
\,{ \pi}\,{q}^{2}}{ \omega^2 \, \left( \! \,{ \Omega}^{2} + 1\,
 \!  \right) \, \left( \! \,{q}^{2} + { \pi}^{2}\, \!  \right) ^{
3}}} + \nonumber \\
 & & {i}\, \left( \! \,{\displaystyle \frac {1}{4}}\,
{\displaystyle \frac {{\it R_1}\,\epsilon\,{ \Omega}^{2}\,
 \left( \! { \Omega}^{2} + 1-\,{L} \, \!  \right) \,{L}\,{ \pi}
\,{q}^{4}}{ \omega^3 \, \left( \! \,{q}^{2} + { 
\pi}^{2}\, \!  \right) ^{5}\, \left( \! \,{ \Omega}^{2} + 1\, \! 
 \right) ^{2}}} + {\displaystyle \frac {1}{2}}\,{\displaystyle 
\frac { \left( \! \,2\,{L}^{2}\,{ \pi}^{2} + { \Omega}^{2}\,{ \pi
}^{2} + {q}^{2}\,{ \Omega}^{2}\, \!  \right) \,{ \pi}\,{q}^{2}}{
 \left( \! \,{ \Omega}^{2} + 1\, \!  \right) \,{ \omega}\,
 \left( \! \,{q}^{2} + { \pi}^{2}\, \!  \right) ^{3}}}\, \! 
 \right), \label{e:deffq}
\\
{\it h_1(q)}&=& - \,{\displaystyle \frac {1}{2}}\,
{\displaystyle \frac {\epsilon\,R_1\,{q}^{2}\, \left( \! \, - { 
\pi}^{2}\,{L}\,{ \Omega}^{2} + {L}\,{q}^{2}\,{ \Omega}^{2} - 2\,{
 \pi}^{2} - { \Omega}^{2}\,{ \pi}^{2} + {q}^{2}\,{ \Omega}^{2}\,
 \!  \right) \,}{{ \Omega}^{2}\,(\,1 + {L}\,)\, \left( 
\! \,{L} - { \Omega}^{2} - 1\, \!  \right) \, \left( \! \,{q}^{2}
 + { \pi}^{2}\, \!  \right) \,{ \pi}}} \nonumber \\
 & & \mbox{} + 2\,{\displaystyle \frac {{ \pi}\, \left( \! \,{ 
\Omega}^{2} + 1\, \!  \right) \, \left( \! \,{q}^{2} + { \pi}^{2}
\, \!  \right) ^{2}\,{L}}{{ \Omega}^{2}\, \left( \! \,{L} - { 
\Omega}^{2} - 1\, \!  \right) }},\label{e:defh1q}
\eea
where $R_1$ and $R_2$ have to be replaced by ${\cal R}$ and $R_0(q)$ via
 ${\cal R}=R_0(q)+\epsilon R_1+\epsilon^2 R_2$. 
The values for  $s$, $s_2$ and $h_4$ obtained by
inserting 
(\ref{e:defaq},\ref{e:deffq},\ref{e:defh1q})
 into (\ref{e:wards},\ref{e:wardh4}) agree with those obtained in
(\ref{e:defs}) and (\ref{e:defh4}) up to terms of order $O(L^2)$ and
$O(\epsilon)$ for $s_2$ and $h_4$, respectively. 
These differences are expected based on the discussion of the
correctness of the $L$-dependence of $s_2$ in appendix \ref{s:appa} and
in view of (\ref{e:wardh4}).

\typeout{}
\typeout{}
\typeout{Kommentar bei A1: Deformation der Wellenzahl: ortsabhaengig}
\typeout{endliche Prandt-Zahl Gleichungen ableiten und endliches L -> Appendix}
\typeout{Laengere Diskussion im Vergleich mit Experiment: dr ist gross, deshalb
dispersiv nicht so wichtig Vorzeichen von di und ci. Beide Effekt in gleiche Richtung}
\typeout{di und ci haben in FSP falsche Vorzeichen.
Luecke: v waechst sublinear
Was passiert wenn beide Vorzeichen von di und ci andersherum sind?
Bilder mit fig.1 etc versehen in xvgr!}
\typeout{}
\typeout{}

\typeout{}
\typeout{Bilder: Oszillatioen durch vertikale Linien andeuten}
\typeout{}

\end{document}